\documentclass[journal]{IEEEtran}

\usepackage{fancyhdr}
\usepackage{amsmath}
\usepackage{amssymb}
\usepackage{graphicx}
\usepackage{booktabs}
\usepackage{amsfonts}
\usepackage{multirow}
\usepackage{algorithm}
\usepackage{algorithmic}
\usepackage{mathrsfs}
\usepackage{bm}
\usepackage{exscale}
\usepackage{relsize}
\usepackage{setspace}
\usepackage{color}
\usepackage{ulem}
\newtheorem{lemma}{\textbf{Lemma}}
\newtheorem{theorem}{\textbf{Theorem}}
\newtheorem{corollary}{\textbf{Corollary}}

\ifCLASSINFOpdf
\else
\fi

\begin{document}

\title{On the Multi-User, Multi-Cell Massive Spatial Modulation Uplink: How Many Antennas for Each User?}

\author{
Longzhuang~He,~\IEEEmembership{Student~Member,~IEEE},
Jintao~Wang,~\IEEEmembership{Senior~Member,~IEEE},
Jian~Song,~\IEEEmembership{Fellow,~IEEE},
and Lajos Hanzo,~\IEEEmembership{Fellow,~IEEE}

\thanks{
Longzhuang He, Jintao Wang and Jian Song are with the Department of Electronic Engineering, Tsinghua University, Beijing, 100084, China (helongzhuang@126.com; \{wangjintao, jsong\}@tsinghua.edu.cn).

L. Hanzo is with the School of Electronics and Computer Science, University of Southampton, Southampton SO17 1BJ, U.K. (lh@ecs.soton.ac.uk).

This work was supported by the Beijing Higher Education Young Elite Teacher Project (Grant No. YETP0101) and the National Natural Science Foundation of China (Grant No. 61471221 and No. 61471219).
}
}

\maketitle
\begin{abstract}
Massive spatial modulation aided multiple-input multiple-output (SM-MIMO) systems have recently been proposed as a novel combination of spatial modulation (SM) and of conventional massive MIMO, where the base station (BS) is equipped with a large number of antennas and simultaneously serves multiple user equipment (UE) that employ SM for their uplink transmission. Since the massive SM-MIMO concept combines the benefits of both the SM and massive MIMO techniques, it has recently attracted substantial research interest. In this paper, we study the achievable uplink spectral efficiency (SE) of a multi-cell massive SM-MIMO system, and derive closed-form expressions to asymptotically lower-bound the SE yielded by two linear BS combining schemes, including maximum ratio (MR) combining and zero forcing (ZF) combining, when a sufficiently large number of BS antennas are equipped. The derivation takes into account the impact of transmitter spatial correlations, of imperfect channel estimations, of user-specific power controls and of different pilot reuse factors. The proposed asymptotic bounds are shown to be tight, even when the scale of BS antennas is limited. The new SE results facilitate a system-level investigation of the optimal number of uplink transmit antennas (TAs) $N$ with respect to SE maximization. Explicitly, we provide theoretical insights on the SE of massive SM-MIMO systems. Furthermore, we demonstrate that massive SM-MIMO systems are capable of outperforming the SE of conventional massive MIMOs relying on single-TA UEs.
\end{abstract}

\IEEEpeerreviewmaketitle

\section{Introduction}
\IEEEPARstart{M}{assive} multiple-input multiple-output (MIMO) systems constitute a promising technique for the next-generation cellular telecommunication networks \cite{rusek2013scaling}-\cite{larsson2014massive}, where the base station (BS) is equipped with a large number of antennas and simultaneously serves numerous single-antenna user equipment (UE). By harnessing the huge diversity and multiplexing gain facilitated by the hundreds of antennas at the BS, the spectral and energy efficiency of massive MIMO is orders of magnitude higher than that of the conventional MIMO systems \cite{ngo2013energy}.

Spatial modulation (SM) is another MIMO technique that was recently proposed for reducing the implementation complexity of conventional MIMO transmitters \cite{di2014spatial}-\cite{he2015infinity}. The conventional MIMO structure of the Vertical Bell Laboratories Layered Space-Time (V-BLAST) scheme \cite{wolniansky1998vblast}, employs the same number of radio frequency (RF) chains as the number of transmit antennas (TAs), which leads to both high power dissipation and to bulky BS design. In SM, however, only one TA is activated for each symbol's transmission, hence the SM transmitter only requires a single RF chain. The single-RF structure of SM significantly reduces both the design complexity and the power consumption, which leads to an improved energy efficiency (EE) \cite{di2014spatial}. Furthermore, in SM, the information is carried both by the index of the active antennas as well as by the transmitted classic amplitude-phase modulation (APM) symbols. Hence a spectral efficiency (SE) gain can also be achieved by SM against single-antenna transmission schemes, albeit at the cost of having no transmit diversity.

SM has been extensively studied in the scenario of point-to-point communications. For instance, \cite{wang2012generalised} proposed a two-stage zero forcing (ZF)-based symbol detector for SM, while \cite{liu2014BPDN} and \cite{yu2012compressed} investigated the application of compressive sensing theory in addressing the symbol detection problem of SM. In \cite{yang2016single}, SM was combined with single carrier modulation for supporting transmission in frequency-selective fading channels, while various transmit pre-coding schemes were studied in \cite{yang2016transmit}-\cite{zhang2016performance}. The information-theoretic capabilities of point-to-point SM systems were investigated in \cite{za2014mianalysis}-\cite{ibrahim2016achievable}. More specifically, in \cite{za2014mianalysis}-\cite{raja2014reduced}, closed-form lower bounds on the mutual information of the classic SM systems were proposed. The channel capacity of SM associated with a large array of antennas was explored in \cite{basnayaka2016massive}, where the authors maximized the mutual information by optimizing the distribution of the channel input. The authors of \cite{ibrahim2016achievable} proposed a general framework for evaluating the achievable rate of SM, in which a Gaussian mixture model was exploited to represent the system's input.


Recently the SM technique was proposed to be combined with massive MIMOs, yielding the novel concept of massive SM-MIMO \cite{narasimhan2014large}-\cite{he2016on}. In contrast to the conventional massive MIMO concept relying on single-TA UEs, massive SM-MIMOs would require multiple TAs at the UEs for uplink transmission. Due to the single-RF structure of SM, both the cost and the design complexity of each UE in massive SM-MIMOs is similar to those in conventional massive MIMOs, while the uplink data rates can be boosted by implicitly conveying extra information via the active antenna's index.

More specifically, in \cite{narasimhan2014large}, a large-scale multi-user SM-MIMO system was proposed along with multi-user detection (MUD) schemes. In \cite{wang2015energy}, an uplink transceiver scheme was proposed for massive SM-MIMO operating in frequency-selective fading channels, while in \cite{wang2015multiuser} low-resolution analog-to-digital convertors (ADCs) were invoked for massive SM-MIMO systems in order to reduce the power consumption at the BS. Furthermore, compressive sensing based MUD schemes were proposed in \cite{rodriguez2015low}, in which the sparsity of the SM signals was exploited to strike a favorable tradeoff between the attained and the complexity imposed performance.

While the above-mentioned research has mainly been focused on improving the MUD performance, the authors of \cite{piya2015spectral} investigated the achievable uplink SE in a multi-cell massive SM-MIMO scenario. However, the scenario of \cite{piya2015spectral} was limited to the case, when all the UEs have the same fading statistics, and the multi-cell interference is also assumed to be identical for all the neighboring cells. Hence the conclusions of \cite{piya2015spectral} cannot be directly extended to a realistic multi-cell environment, where the fading statistics are dependent on the user-specific locations and where the multi-cell interference is different in the neighboring cells. Moreover, the uplink spectral efficiency as well as the optimal number of UE TAs in massive SM-MIMO systems were investigated in \cite{he2016on}, which is, however, limited to a single-cell scenario where the BS only uses maximum ratio (MR) combining for MUD. More importantly, the impact of pilot reuse and of power control was not explored in \cite{piya2015spectral} and \cite{he2016on}, which prevents their applications in a more generalized context of massive SM-MIMO.

Against this background, the novel contributions of this paper are summarized as follows.
\begin{itemize}
  \item A generalised theoretical framework is proposed for the SE analysis of massive SM-MIMO systems relying on realistic channel fading and inter-cell interference, where the fading and interfering channels' coefficients are correlated with the UEs' random distribution and pilot-reuse schemes. The impact of both the user power control and of the TAs' spatial correlations is also accounted for in our work.

  \item At an asymptotically large number of BS antennas, lower bounds are derived for quantifying the achievable uplink SE in the case of a \textit{fixed} geographic UE-distribution, which are then extended to the general case of a \textit{random} UE distribution. The proposed SE expressions are shown to be tight for various system parameters, even when the scale of BS antennas is limited.

  \item Based on our new theoretical framework, a heuristic system-level optimization is carried out for finding the optimal number of TAs for each UE, which constitutes the most influential parameter of massive SM-MIMO systems. To the best of our knowledge, this issue has not been addressed for multi-cell massive SM-MIMO systems. Finally massive SM-MIMOs are shown to be capable of outperforming the conventional massive MIMOs relying on single-TA UEs.
\end{itemize}

The organization of this paper is summarized as follows. Section \uppercase\expandafter{\romannumeral2} introduces the general model of our multi-cell massive SM-MIMO system along with our uplink pilot-based channel estimation (CE) scheme. Section \uppercase\expandafter{\romannumeral3} introduces our theoretical framework conceived for the SE analysis under the assumption of fixed UE locations. Section \uppercase\expandafter{\romannumeral4} generalizes the results of Section \uppercase\expandafter{\romannumeral3} to the case of random UE locations. Section \uppercase\expandafter{\romannumeral5} provides our simulation results, where we seek to optimize the number of UE antennas for the purpose of SE maximization. Section \uppercase\expandafter{\romannumeral6} concludes this paper and briefly introduces our future work.

\textit{Notations}: in this paper, $\mathcal{CN}(\bm\mu, \bm\Sigma)$ denotes a circularly symmetric complex-valued multi-variate Gaussian distribution with $\bm\mu$ and $\bm\Sigma$ being its mean and covariance, respectively, while $\mathcal{CN}(\mathbf{x}; \bm\mu, \bm\Sigma)$ denotes the probability density function (PDF) of a random vector $\mathbf{x} \sim \mathcal{CN}(\bm\mu, \bm\Sigma)$. $\mathbf{M}(i, j)$ is used to denote the $(i; j)$ component of a matrix $\mathbf{M}$, and $\text{diag}\{\mathbf{A}_k\}_{k=1}^K$ denotes a diagonal matrix with $\mathbf{A}_k$ being its $k$-th diagonal sub-matrix. $\{0, 1\}^N$ denotes an integer vector composed of $N$ elements selected from $0$ and $1$. $\mathbf{I}_N$ denotes an $N$-dimensional identity matrix.

\section{Multi-Cell Massive SM-MIMO System Model}
\begin{figure*}
\center{\includegraphics[width=0.75\linewidth]{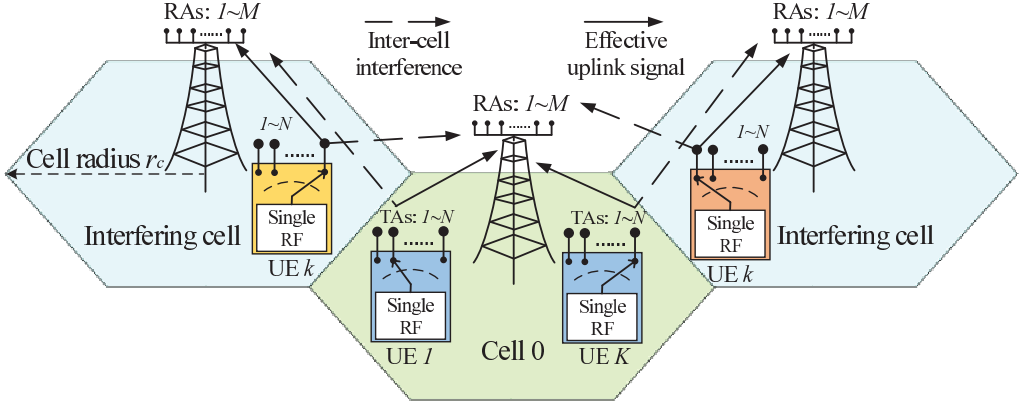}}
\caption{Multi-cell massive SM-MIMO uplink model, where $K$ UEs are simultaneously scheduled by a massive-antenna BS at the center of a cell. Each UE is equipped with $N$ TAs and uses SM for its uplink transmission.}
\label{Fig_MultiCellUplink}
\end{figure*}

\subsection{System Model}
We consider the multi-cell uplink massive SM-MIMO system of Fig.\ref{Fig_MultiCellUplink}, where the BS of each cell is equipped with $M$ receive antennas (RAs) and simultaneously serves $K$ UEs. Each UE is equipped with $N$ TAs and uses SM for its uplink transmission. The BS is placed at the center of a hexagonal cell with radius $r_\text{c}$, and we assume that the parameters $M$, $K$ and $N$ are the same for all the cells.

Furthermore, the uplink transmission is divided into several time-domain frames consisting of $T_\text{c}$ seconds and the SM symbols are transmitted at $1/T_\text{s}$ symbols per second, which leaves room for $T=T_\text{c}/T_\text{s}$ transmitted symbols in each frame. Moreover, the frame duration $T_\text{c}$ is designed to be shorter or equal to the channel's coherence time, hence all the channel impulse responses (CIRs) can be assumed to be time-invariant within each frame. We let $\mathbf{h}_{jkn} \in \mathbb{C}_{M \times 1}$ denote the CIR of the link spanning from the $n$-th TA of UE $k$ in cell $j$ to the BS in cell $0$, which can, according to \cite{bjornson2016massive} and \cite{ngo2013energy}, be modeled as a zero-mean circularly symmetric complex-valued Gaussian random vector, i.e.
\begin{equation}
  \mathbf{h}_{jkn} \sim \mathcal{CN}(\mathbf{0}, \beta_{0jk}\mathbf{I}_M),
\end{equation}
where $\beta_{ljk} > 0$ characterizes the large-scale attenuation between UE $k$ of cell $j$ and BS $l$. Since the TAs of each UE are usually compactly placed due to the limited dimensions of the UE, it is reasonable to assume that the large-scale attenuations of different TAs of a specific UE are the same, hence $\beta_{ljk}$ is independent of the TA index $n$.

Let $\mathbf{H}_{jk} \in \mathbb{C}_{M \times N}$ denote the spatially-correlated MIMO channel matrix between UE $k$ of cell $j$ and BS $0$, we thus have:
\begin{equation}
  \mathbf{H}_{jk} \triangleq [\mathbf{h}_{jk1}, \ldots, \mathbf{h}_{jkN}] = \sqrt{\beta_{0jk}} \tilde{\mathbf{H}}_{jk} \mathbf{R}_\text{t}^{\frac{1}{2}},
  \label{Hjk}
\end{equation}
where $\tilde{\mathbf{H}}_{jk} \in \mathbb{C}_{M \times N}$ is composed of i.i.d. $\mathcal{CN}(0, 1)$ random elements, and $\mathbf{R}_\text{t} \in \mathbb{R}_{N \times N}$ is the correlation matrix at the transmitter side. As in \cite{kuo2012compressive}, we assume the TAs of each UE to form a uniformly-spaced linear array, which leads to having correlation coefficients governed by Jakes' model, i.e.
\begin{equation}
  \mathbf{R}_\text{t}(i, j) = J_0\left(\frac{2 \pi d_\text{s} |i-j|}{\lambda}\right),
  \label{CorrModel}
\end{equation}
where $d_\text{s}$ is the minimum distance between the adjacent TAs of each UE, $\lambda$ is the carrier's wavelength and $J_0(\cdot)$ denotes a zero-order Bessel function of first kind. It is worth noting that the dimension of each UE is much more limited compared to the BS, hence we assume having no spatial correlation at the BS side and focus our attention on the impact of the uplink TA correlations.

\subsection{Uplink Pilot-Based Channel Estimation}
\begin{figure}
\center{\includegraphics[width=0.95\linewidth]{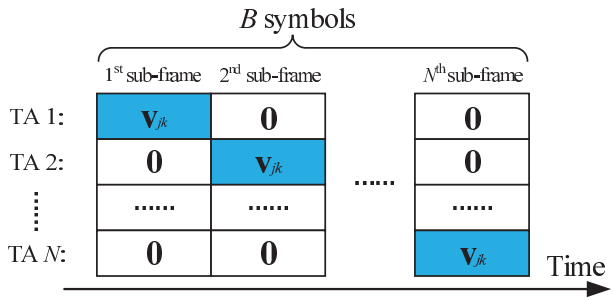}}
\caption{Pilot signaling scheme for UE $k$ in cell $j$, where $\mathbf{v}_{jk} \in \mathbb{C}_{B/N\times 1}$ is a time-domain-orthogonal sequence, and $\mathbf{0}$ denotes de-activating the corresponding antenna.}
\label{Fig_PilotSignaling}
\end{figure}

\begin{figure}
\center{\includegraphics[width=0.95\linewidth]{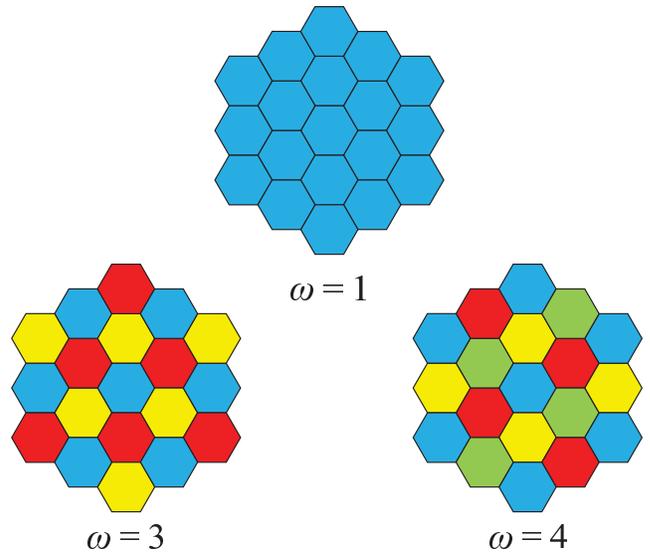}}
\caption{Pilot reuse pattern for $\omega=1$, $3$ and $4$, where the same pilot signals are reused in the cells that are filled with the same color.}
\label{Fig_PilotReusePattern}
\end{figure}

CE is performed at the BS using the received uplink pilots. As stated in \cite{ngo2013energy}, during the uplink transmission, $B \ge 1$ out of the $T$ symbols in each frame are reserved for pilot transmission, and CE is performed at the BS side. The uplink pilots of cell $j$ are designed as seen in Fig.\ref{Fig_PilotSignaling}, where the $B$ symbols are divided into $N$ sub-frames, and TA $n$ is exclusively activated by UE $k$ in the $n$-th sub-frame to transmit a time-domain-orthogonal sequence $\mathbf{v}_{jk} \in \mathbb{C}_{(B/N) \times 1}$. By activating the TAs in a one-by-one manner as in Fig.\ref{Fig_PilotSignaling}, the number of RF chains required is reduced to $1$, which is eminently suitable for the single-RF-chain UEs using SM.

We study the multi-cell \textit{fractional pilot reuse} regime of \cite{bjornson2016massive} and \cite{bjornson2014optimizing}, where only a fraction $1/\omega$ of the cells use the same pilot symbols as cell $0$, and $\omega \ge 1$ is an integer referred to as the \textit{pilot reuse factor}. In this paper, we use $\Phi$ to represent all the neighboring cells, and $\Phi'$ to represent the set of cells that use the same pilots as cell $0$. The pilot reuse patterns associated with $\omega=1$, $3$ and $4$ are depicted in Fig.\ref{Fig_PilotReusePattern}, where the same pilots are reused in the cells filled with the same color. Assuming that the cells in $\Phi'_0 \triangleq \{0, \Phi'\}$ assign the same uplink pilot signals to the $k$-th UE, we thus have:
\begin{equation}
  \mathbf{v}_{0k}^H \mathbf{v}_{jm} =
  \begin{cases}
    \displaystyle \frac{B}{N}, & \text{if}\,\, j \in \Phi'_0 \,\, \text{and} \,\, m = k, \\
    0, & \text{otherwise}, \\
  \end{cases}
  \label{PilotRequirement}
\end{equation}
where $j \in \Phi_0$, and $\Phi_0 \triangleq \{0, \Phi\}$ denote all the cells that are considered in our analysis. In order that the orthogonality in (\ref{PilotRequirement}) can be satisfied, it is required that we have $B = \omega NK$, and the pilot signals $\mathbf{v}_{jk}$ can hence be chosen from the columns of a $(B/N = \omega K)$-dimensional discrete Fourier transform (DFT) matrix. Therefore the set of symbols received by BS $0$ in the $n$-th sub-frame during the pilot transmission is denoted by $\mathbf{Y}_{\text{p}, n} \in \mathbb{C}_{M \times \omega K}$ and given by:
\begin{equation}
  \mathbf{Y}_{\text{p}, n} = \displaystyle\sum_{j \in \Phi_0} \sum_{k=1}^K \mathbf{h}_{jkn} \mathbf{v}_{jk}^H \sqrt{P_{jk}} + \mathbf{N}_{\text{p}, n},
  \label{UplinkReceivedPilots}
\end{equation}
where $P_{jk} > 0$ is the transmit power of UE $k$ in cell $j$ and $\mathbf{N}_{\text{p}, n}$ is composed of i.i.d. $\mathcal{CN}(0, \sigma_\text{N}^2)$ elements, while $\sigma_\text{N}^2 > 0$ is the power of the additive white Gaussian noise (AWGN) on the RAs of each BS.


In contrast to \cite{piya2015spectral} and \cite{he2016on}, we consider a channel-statistics-aware user power control strategy, as in \cite{bjornson2016massive}, where we assume that the channel statistics $\beta_{jjk}$ are slowly varying over time, and they are known to both UE $k$ and to the BS of cell $j$. The power control strategy is therefore designed as $P_{jk} = P_\text{u} / \beta_{jjk}$, where $P_\text{u} > 0$ represents the \textit{effective received power} of each UE in the cell. This UE-specific power control policy adjusts the effective channel gains for all the UEs in cell $0$ to $E\{P_{0k}\|\mathbf{h}_{0kn}\|^2\} = M P_\text{u}$, which has the benefit of maintaining a uniform service quality for all the UEs in this cell. In practice, the parameter $P_\text{u}$ should be carefully selected so that the power of the cell-edge UEs does not exceed the power limit. However, as stated in \cite{bjornson2016massive}, the performance of massive MIMO will not be severly degraded at a low signal-to-noise ratio (SNR), hence it will not be a critical limitation in deploying our system. In practice, we might also occasionally have to drop the UEs having extremely bad channel conditions to implement this power control policy, as stated in \cite{bjornson2016massive}.

The ZF based channel estimation of $\mathbf{h}_{0kn}$ is hence formulated as:
\begin{equation}
    \hat{\mathbf{h}}_{0kn} = \displaystyle\frac{\mathbf{Y}_{\text{p}, n} \mathbf{v}_{0k}}{\omega K \sqrt{P_{0k}}} = \displaystyle \mathbf{h}_{0kn} + \sum_{j \in \Phi'} \mathbf{h}_{jkn} \sqrt{\frac{\beta_{00k}}{\beta_{jjk}}} + \mathbf{w}_{kn},
    \label{Esti_h_0kn}
\end{equation}
in which the second equality holds according to the orthogonality requirement encapsulated in (\ref{PilotRequirement}), and $\mathbf{w}_{kn} \sim \mathcal{CN}(\mathbf{0}, \frac{\sigma_\text{N}^2 \beta_{00k}}{\omega K P_\text{u}} \mathbf{I}_M )$. Observe that the estimate of $\mathbf{h}_{0kn}$ is inevitably affected by the interference imposed by the pilot-reuse cells in $\Phi'$, which results in the so-called \textit{pilot contamination} phenomenon and deteriorates the achievable uplink SE. The imperfect channel estimate in (\ref{Esti_h_0kn}) will be utilized in all the derivations and simulations of this paper.

\section{Uplink SE Analysis for Fixed UE Locations}
In this section we will provide a general theoretical framework for analyzing the achievable uplink SE in a massive SM-MIMO system having fixed UE locations, which is later invoked for quantifying the SE achieved by maximum ratio (MR) and ZF combining. In this section, the only randomly distributed elements are the fading channels, the thermal noise and the transmitted uplink data symbols.

\subsection{Theoretical Framework}
Let $x_{jkn} \in \mathbb{C}_{1\times 1}$ denote the symbol transmitted on TA $n$ of UE $k$ in cell $j$ with $E\{|x_{jkn}|^2\}=1$. The signals received by the BS's RAs in cell $0$ during the uplink transmission can hence be denoted by $\mathbf{y}_\text{u} \in \mathbb{C}_{M \times 1}$ and given by:
\begin{equation}
    \mathbf{y}_\text{u} = \displaystyle \sum_{j \in \Phi_0} \sum_{k=1}^K \sum_{n=1}^N x_{jkn} \gamma_{jkn} \mathbf{h}_{jkn} \sqrt{P_{jk}} + \mathbf{n}_\text{u},
    \label{ReceivedUplinkData}
\end{equation}
where $\mathbf{n}_\text{u} \sim \mathcal{CN}(\mathbf{0}, \sigma_\text{N}^2 \mathbf{I}_M)$ denotes the AWGN received by the BS's RAs. Note that the subscript ``u'' in $\mathbf{y}_\text{u}$ and $\mathbf{n}_\text{u}$ represents the word ``uplink''. Furthermore, $\gamma_{jkn} \in \{0, 1\}$ is a binary random variable characterizing the activity of the $n$-th TA of UE $k$ in cell $j$, where $\gamma_{jkn} = 0$ and $1$ denotes a de-activated and an activated antenna, respectively. According to the SM principle, it is required that $\sum_{n=1}^N \gamma_{jkn} = 1$, and $\gamma_{jkn}=1$ with a probability of $1/N$. Equation (\ref{ReceivedUplinkData}) follows from modeling each specific TA as an independent ``UE''. The corresponding transmitted symbol is denoted by $x_{jkn}\gamma_{jkn} \sqrt{P_{jk}}$, and the received signal is thus obtained by overlapping them as in the classic model of multi-user massive MIMO systems \cite{bjornson2016massive}\cite{ngo2013energy}.

Moreover, we assume that BS $0$ uses a vector $\mathbf{g}_{0kn}$ for linearly amplifying the signal power of the $n$-th TA of UE $k$ and rejecting the interference from other UEs. According to the theory in \cite{bjornson2016massive}, we formulate Lemma \ref{lemma1} for quantifying the post-processing signal-to-interference-and-noise ratio (SINR) of the symbol transmitted by TA $n$ of UE $k$ in cell $0$.

\begin{figure*}[t]
\normalsize
\begin{equation}
\text{SINR}_{kn} = \displaystyle \frac{\displaystyle\frac{P_{0k}}{N} \left|E_{\mathbf{h}}\left\{\mathbf{g}_{0kn}^H \mathbf{h}_{0kn}\right\}\right|^2}{\displaystyle \sum_{j' \in \Phi_0} \sum_{k'=1}^K \sum_{n'=1}^N \frac{P_{j'k'}}{N} E_{\mathbf{h}} \left\{ \left| \mathbf{g}_{0kn}^H \mathbf{h}_{j'k'n'}\right|^2\right\} - \displaystyle\frac{P_{0k}}{N} \left|E_{\mathbf{h}}\left\{\mathbf{g}_{0kn}^H \mathbf{h}_{0kn}\right\}\right|^2 + \sigma_\text{N}^2 E_{\mathbf{h}} \left\{ \left\| \mathbf{g}_{0kn} \right\|^2 \right\}}.
\label{SINRBound}
\end{equation}
\hrulefill
\end{figure*}

\begin{lemma}
  By performing linear combining using $\mathbf{g}_{0kn}$, the SINR of the symbol transmitted by the $n$-th TA of UE $k$ in cell $0$ for fixed UE locations is lower-bounded by (\ref{SINRBound}), where $E_\mathbf{h}\{\cdot\}$ denotes taking the expectations over the random channel realizations.
  \label{lemma1}
\end{lemma}


\begin{IEEEproof}
  The proof of Lemma \ref{lemma1} follows from a direct application of \cite[Lemma~2]{bjornson2016massive}, in which the interference and noise terms are handled as worst-case additive Gaussian noise. We refer the interested readers to \cite{bjornson2016massive} for more theoretical details.
\end{IEEEproof}

Aided by Lemma \ref{lemma1}, the post-processing signal $\mathbf{y}_{\text{post}, k} \in \mathbb{C}_{N \times 1}$ of the $k$-th UE can hence be equivalently modeled as the output of an additive Gaussian noise channel, i.e. we have:
\begin{equation}
  \mathbf{y}_{\text{post}, k} = \mathbf{x}_{\text{SM}, k} + \mathbf{w}_{\text{eff}, k},
  \label{EqChnModel1}
\end{equation}
where $\mathbf{x}_{\text{SM}, k} \in \mathbb{C}_{N \times 1}$ is the transmitted SM signal having a covariance of $E\{\mathbf{x}_{\text{SM}, k} \mathbf{x}_{\text{SM}, k}^H\} = \mathbf{I}_N$, while $\mathbf{w}_{\text{eff}, k}$ is the zero-mean circularly symmetric complex-valued additive Gaussian noise having a covariance of:
\begin{equation}
  E\left\{ \mathbf{w}_{\text{eff}, k} \mathbf{w}_{\text{eff}, k}^H\right\} = \text{diag}\left\{ \text{SINR}^{-1}_{k1}, \ldots, \text{SINR}^{-1}_{kN} \right\},
  \label{EqChnModel2}
\end{equation}
where $\text{SINR}_{kn}$ has been given by (\ref{SINRBound}). The achievable SE of UE $k$ associated with fixed UE locations, i.e. with $R_k^\text{fixed}$, is hence formulated as the mutual information between $\mathbf{y}_{\text{post}, k}$ and $\mathbf{x}_{\text{SM}, k}$ expressed as:
\begin{equation}
  R_k^\text{fixed} = \displaystyle \frac{T - B}{T} I(\mathbf{y}_{\text{post}, k}; \mathbf{x}_{\text{SM}, k}) \,\,[\text{bits/s/Hz}],
  \label{Rfixed}
\end{equation}
where $(T-B)/T$ is the normalized effective data transmission time. The calculation of the mutual information term in (\ref{Rfixed}) relies on numerical integrations and lacks a tractable closed-form formula, hence we propose Lemma \ref{lemma2} to lower-bound $R_k^\text{fixed}$ by $R_k^\text{fixed,LB}$ in a tractable form.

\begin{lemma}
  In a massive SM-MIMO system having fixed UE locations, the achievable rate of the $k$-th UE is lower-bounded by $R_k^\text{fixed, LB}$ given by
  \begin{equation}
  \arraycolsep=1.0pt\def\arraystretch{1.3}
  \begin{array}{rcl}
    && R_k^\text{fixed,LB} = \displaystyle\frac{T-B}{T} \left[ \log_2(1 + N \sigma_{k}^{-2}) + \log_2 N + \,\text{...}\right. \\
    && \left. P_\text{c} \log_2 P_\text{c} + (1-P_\text{c})\log_2\left(\displaystyle \frac{1-P_\text{c}}{N-1}\right) \right] \,\,[\text{bits/s/Hz}],
  \end{array}
  \label{Rfixed_LB}
  \end{equation}
  where $\sigma_k^2 = 1/N \sum_{n=1}^N\text{SINR}_{kn}^{-1}$, and $P_\text{c}$ is the probability of a correct TA detection given by:
  \begin{equation}
    P_\text{c} = \displaystyle\frac{\displaystyle \sum_{r=0}^{N-2} \binom{N-2}{r}(-1)^r \left(r + \displaystyle \frac{N + 2\sigma_k^2}{N + \sigma_k^2}\right)^{-1}}{\displaystyle \sum_{r=0}^{N-2} \binom{N-2}{r}(-1)^r \left(r + 1\right)^{-1}}.
    \label{PcDef}
  \end{equation}
  \label{lemma2}
\end{lemma}

\begin{IEEEproof}
  The proof is provided in the Appendix.
\end{IEEEproof}

In the next subsection, we will extend the SINR formula of (\ref{SINRBound}) to MR and ZF combining, and formulate the asymptotic expressions of $\text{SINR}_{kn}^\text{MR}$ and $\text{SINR}_{kn}^\text{ZF}$ in the context of large-scale MIMO.

\subsection{Asymptotic SE Lower Bounds Achieved by Linear Combining Schemes}
We let $\mathbf{g}_{0kn}^\text{MR} \in \mathbb{C}_{M \times 1}$ and $\mathbf{g}_{0kn}^\text{ZF} \in \mathbb{C}_{M \times 1}$ denote the linear combining vector of MR and ZF with respect to the $n$-th TA of UE $k$ in cell $0$, respectively. For MR combining we have $\mathbf{g}_{0kn}^\text{MR} \triangleq \hat{\mathbf{h}}_{0kn}$, where $\hat{\mathbf{h}}_{0kn}$ is given by (\ref{Esti_h_0kn}). In order to derive $\mathbf{g}_{0kn}^\text{ZF}$, we first let:
\begin{equation}
  \mathbf{H}_j \triangleq \left[ \mathbf{H}_{j1}, \ldots, \mathbf{H}_{jK} \right] \in \mathbb{C}_{M \times NK},
  \label{Hj}
\end{equation}
where $\mathbf{H}_{jk}$ is given by (\ref{Hjk}). Therefore, based on (\ref{Esti_h_0kn}), the collective representation of $\hat{\mathbf{h}}_{0kn}$ with $1\le k \le K$ and $1 \le n \le N$ is denoted by $\hat{\mathbf{H}}_0 \in \mathbb{C}_{M \times NK}$ and given by:
\begin{equation}
\hat{\mathbf{H}}_0 = [\hat{\mathbf{H}}_{01}, \ldots, \hat{\mathbf{H}}_{0K}] = \displaystyle \sum_{j \in \Phi'_0} \mathbf{H}_j \mathbf{A}_j^{\frac{1}{2}} + \mathbf{W},
\label{Hhat0}
\end{equation}
where $\hat{\mathbf{H}}_{0k} \in \mathbb{C}_{M \times N}$ is given by $\hat{\mathbf{H}}_{0k} = [\hat{\mathbf{h}}_{0k1}, \ldots, \hat{\mathbf{h}}_{0kN}]$, and $\mathbf{A}_j \in \mathbb{C}_{NK \times NK}$ is a diagonal matrix formulated as:
\begin{equation}
  \mathbf{A}_j = \displaystyle \text{diag}\left\{\frac{\beta_{00k}}{\beta_{jjk}}\mathbf{I}_N\right\}_{k=1}^K,
  \label{Aj}
\end{equation}
according to the expression of $\hat{\mathbf{h}}_{0kn}$ in (\ref{Esti_h_0kn}). The noise term $\mathbf{W}$ in (\ref{Hhat0}) is composed of i.i.d. complex-valued Gaussian random variables, of which the $[(k-1)N + n]$-th column is distributed according to $\mathcal{CN}(\mathbf{0}, \frac{\sigma_\text{N}^2 \beta_{00k}}{\omega K P_\text{u}}\mathbf{I}_M)$. With the expression in (\ref{Hhat0}), $\mathbf{g}_{0kn}^\text{ZF}$ can hence be represented as the $[(k-1)N+n]$-th column of matrix $( \hat{\mathbf{H}}_0^\dagger )^H$, where we have $\hat{\mathbf{H}}_0^\dagger = (\hat{\mathbf{H}}_0^H \hat{\mathbf{H}}_0)^{-1} \hat{\mathbf{H}}_0^H$. Based on the expressions of $\mathbf{g}_{0kn}^\text{MR}$ and $\mathbf{g}_{0kn}^\text{ZF}$, we arrive at Theorem \ref{theorem1} for asymptotically lower-bounding the achievable SE for MR and ZF combining.


\begin{figure*}[t]
\normalsize
\begin{equation}
\frac{1}{\text{SINR}_{kn}^\text{MR}} = (1 + \epsilon_\text{s}) \displaystyle \sum_{j \in \Phi'}\mu_{jk}^2 + \epsilon_\text{s} + \frac{N}{M} \left( \frac{\sigma_\text{N}^2}{\omega K P_\text{u}} + \sum_{j \in \Phi'_0} \mu_{jk} \right) \left( \frac{\sigma_\text{N}^2}{P_\text{u}} + \sum_{j' \in \Phi_0} \sum_{k'=1}^K \mu_{j'k'}\right).
\label{inverseSINR_MR}
\end{equation}

\begin{equation}
\frac{1}{\text{SINR}_{kn}^\text{ZF}} = \displaystyle\sum_{j\in\Phi'}\mu_{jk}^2 + \displaystyle\frac{r_n N}{M-NK} \displaystyle\sum_{j\in\Phi'_0}\mu_{jk}  \left( \sum_{j' \in \Phi'_0}\mu_{j'k} + \sum_{j' \in \Phi'_0}\sum_{\substack{k'=1 \\ k' \ne k}}^K \theta_\omega \mu_{j'k'} + \sum_{j'\notin\Phi'_0}\sum_{k'=1}^K \mu_{j'k'} + \frac{\sigma_\text{N}^2}{P_\text{u}}  \right).
\label{inverseSINR_ZF}
\end{equation}
\hrulefill
\end{figure*}

\begin{theorem}
At an asymptotically large number of BS antennas, the achievable SE of UE $k$ in cell $0$ using MR and ZF combining can be lower-bounded by replacing $\sigma_k^2$ in Lemma \ref{lemma2} with $1/N\sum_{n=1}^N (\text{SINR}_{kn}^\text{MR})^{-1}$ and $1/N\sum_{n=1}^N (\text{SINR}_{kn}^\text{ZF})^{-1}$, respectively. The expressions of $1/\text{SINR}_{kn}^\text{MR}$ and $1/\text{SINR}_{kn}^\text{ZF}$ are given by (\ref{inverseSINR_MR}) and (\ref{inverseSINR_ZF}) in conjunction with:
\begin{equation}
\epsilon_\text{s} = \displaystyle\sum_{n=2}^N \left[\mathbf{R}_\text{t}(1, n)\right]^2, \,\, \mu_{jk} = \displaystyle \frac{\beta_{0jk}}{\beta_{jjk}}, \,\, r_n = \mathbf{R}_\text{t}^{-1}(n, n),
\end{equation}
and $\theta_\omega$ is a scaling factor, which is set to $0.2$ and $0.01$ when $\omega=1$ and $\omega>1$, respectively.
  \label{theorem1}
\end{theorem}

\begin{IEEEproof}
  The proof is provided in the Appendix.
\end{IEEEproof}

In the next subsection, the expressions (\ref{inverseSINR_MR}) and (\ref{inverseSINR_ZF}) are shown to be simplified and become dependent on only a few parameters, when $M$ tends to infinity.

\subsection{Asymptotic Analysis For Large-Scale BS Antennas}
By increasing $M$ without limit, the following corollary can be immediately formulated:
\begin{corollary}
Let $\Phi'$ denote the cells that use the same pilots as cell $0$. Then the reciprocal of the SINR for MR and ZF combining derived for fixed UE locations converges to the following limits, when $M \rightarrow \infty$:
\begin{equation}
\arraycolsep=1.0pt\def\arraystretch{1.3}
\begin{array}{rcl}
\displaystyle \frac{1}{\text{SINR}_{kn}^\text{MR}} &\rightarrow& (1+\epsilon_\text{s}) \displaystyle \sum_{j \in \Phi'} \mu_{jk}^2 + \epsilon_\text{s}, \\
\displaystyle \frac{1}{\text{SINR}_{kn}^\text{ZF}} &\rightarrow& \displaystyle \sum_{j \in \Phi'} \mu_{jk}^2,
\end{array}
\label{AsympLimFixedUE}
\end{equation}
where $\epsilon_\text{s} > 0$ and $\mu_{jk} > 0$ are given in Theorem \ref{theorem1}.
\label{corollary1}
\end{corollary}

From Corollary \ref{corollary1}, the effects of pilot contamination and transmit spatial correlation imposed on the effective SINR become explicit. More specifically, for MR combining, the TA's spatial correlation term $\epsilon_\text{s}$ in (\ref{AsympLimFixedUE}) directly increases the SINR's reciprocal, hence degrading the SINR. Furthermore, the TA's correlation term $\epsilon_\text{s}$ also \textit{amplifies} the pilot contamination term $\sum_{j\in\Phi'}\mu_{jk}^2$ by the factor of $(1+\epsilon_\text{s})$. This result encourages us to carefully adjust the spacing of UE antennas, so that the correlation term $\epsilon_\text{s}$ is minimized, which is equivalent to finding the optimal TA spacing $d_\text{s}^*$ formulated as:
\begin{equation}
  d_\text{s}^* = \displaystyle \arg \min_{0 < d_\text{s} \le d_\text{s}^\text{m}} \sum_{n=2}^N \left[J_0\left( \frac{2\pi d_\text{s} \left| n-1 \right|}{\lambda} \right)\right]^2,
  \label{TranAntSpOpt}
\end{equation}
where $d_\text{s}^\text{m} \triangleq D_\text{m}/(N-1)$ denotes the maximum possible antenna spacing, and $D_\text{m}$ represents the device's dimension of each UE.

In contrast to MR combining, when ZF combining is invoked, Corollary \ref{corollary1} shows that the TA correlation coefficient $\epsilon_\text{s}$ is not involved at all when $M$ is increased without limit. However, it is worth noting that the array gain of the large-scale BS antennas is only $(M-NK)/(Nr_n)$ for ZF combining, which is less than the gain $M/N$ in the MR combining scheme. A direct impact of this antenna-array gain reduction of ZF is that the UE may not benefit as much from SM for ZF, as for MR combining at a specific $M$ value.

\subsection{Bound Tightness}
In this subsection we report on our numerical simulations to validate the tightness of the proposed asymptotic bounds in Theorem \ref{theorem1}. We will compare the theoretical results of Theorem \ref{theorem1} against the simulated SE $R_k^\text{fixed}$ in (\ref{Rfixed}). To compute $R_k^\text{fixed}$ in (\ref{Rfixed}), numerical integration-based mutual information calculation is performed according to (\ref{Rfixed}) and the expectation terms in the SINR expression of (\ref{SINRBound}) are calculated via Monte-Carlo simulations over $10,000$ channel realizations.

\begin{table}
\small
\caption{Simulation Parameters}
\newcommand{\tabincell}[2]{\begin{tabular}{@{}#1@{}}#2\end{tabular}}
\centering
\renewcommand\arraystretch{1.2}
\begin{tabular}{c|l|l}
\hline\hline
Symbols & Specifications & Typical Values \\\hline
$M$     & Number of BS's RAs            & $512$ \\\hline
$N$     & Number of TAs on each UE      & $2$ \\\hline
$K$     & Number of UEs in each cell    & $10$ \\\hline
$B$     & \tabincell{l}{Number of symbols reserved\\for uplink pilot transmission} & $\omega NK$ \\\hline
$T$     & \tabincell{l}{Number of symbols\\ transmitted per frame} & $1000$ \\\hline
$\omega$ & Pilot reuse factor & $3$ \\\hline
$P_\text{u} / \sigma_\text{N}^2$ & Effective SNR of each UE & $10$ dB \\\hline
$D_\text{m}$ & Device size & $100$ mm \\\hline
$\lambda$ & Carrier's wavelength & $60$ mm \\\hline
$r_\text{c}$ & Cell radius & $500$ m \\\hline
$r_\text{min}$ & \tabincell{l}{Minimum distance between \\the UEs and the BS} & $0.1 r_\text{c}$  \\\hline
$\alpha$ & Path loss exponent & $3.7$ \\\hline
\hline
\end{tabular}
\label{TABLESimu}
\end{table}


Moreover, we consider a $19$-cell network model with cell $0$ at the center, as depicted in Fig.\ref{Fig_PilotReusePattern}. The $K$ UEs in cell $j$ ($j \in \Phi_0$) are uniformly placed at a $275$-meter distance from the center of cell $j$, while the cell radius is configured as $r_\text{c} = 500 $m. Similar to \cite{bjornson2016massive}, the large-scale attenuation $\beta_{ljk}$ is given by $(d_{ljk}/r_\text{min})^{-\alpha}$, where $d_{ljk}$ is the distance of UE $k$ in cell $j$ to BS $l$, $r_\text{min} = 0.1 r_\text{c}$ is the minimum distance between each UE and its serving BS, and $\alpha$ is the path loss exponent. Note that according to (\ref{SINRBound}), (\ref{inverseSINR_MR}) and (\ref{inverseSINR_ZF}), the SINR is only affected by $P_\text{u}/\sigma_\text{N}^2$, we hence only have to specify the ratio $P_\text{u}/\sigma_\text{N}^2$ for our simulations. For convenience, all the simulation parameters have been summarized in Table \ref{TABLESimu} along with their specifications and typical values. For all the simulations in this paper, the corresponding parameters are based on Table \ref{TABLESimu}, unless stated otherwise.

\begin{figure}
\center{\includegraphics[width=0.95\linewidth]{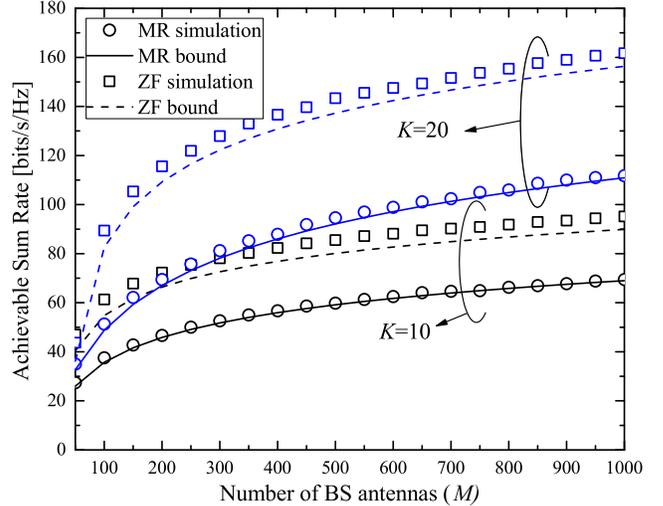}}
\caption{Achievable sum rate with respect to various $M$ and $K$. All the other parameters have been specified in Table \ref{TABLESimu}. The lines are computed based on Theorem \ref{theorem1} while the markers are obtained via (\ref{Rfixed}). }
\label{Fig_BoundTightnessFixed}
\end{figure}

In Fig.\ref{Fig_BoundTightnessFixed}, the \textit{per-cell sum rate}, which is defined as the sum of the $K$ UEs' achievable uplink SE in cell $0$, are depicted for various $M$ and $K$ values. As seen from the figure, the simulation results agree well with the asymptotic bound for MR combining, while a modest deviation is observed for ZF combining due to the independence assumptions applied in the derivations. In a nutshell, the proposed asymptotic bounds are relatively tight, even when a limited $M$ value is used. Let us now move on to investigate the achievable SE for random UE locations based on the results of this section.

\section{Uplink SE Analysis for Random UE Locations}

\subsection{Asymptotic SE Bounds Achieved by Linear Combining Schemes}
Let $\mathbf{z}_{jk} \in \mathbb{C}_{2\times1}$ denote the $2$-dimensional location of UE $k$ in cell $j$. Then an asymptotic SE bound averaged over the UE distribution across the cell is given by Lemma \ref{lemma3}:

\begin{lemma}
  An asymptotic SE bound for UE $k$ in cell $0$ associated with random UE locations is given by:
  \begin{equation}
  \arraycolsep=1.0pt\def\arraystretch{1.3}
  \begin{array}{rcl}
  && R_k^\text{random} = \displaystyle E_\mathbf{z}\left\{ R_k^{\text{fixed,LB}} \right\} \\
  && = \displaystyle \frac{T-B}{T} E_\mathbf{z}\left\{ \log_2(1 + N \sigma_k^{-2}) + \log_2 N + \,\text{...} \right. \\
  && \left. P_\text{c} \log_2 P_\text{c} + (1 - P_\text{c}) \log_2\left( \displaystyle\frac{1-P_\text{c}}{N-1}\right)  \right\} \,\, \left[\text{bits/s/Hz}\right],
  \end{array}
  \label{Rrandom}
  \end{equation}
  where $R_k^{\text{fixed,LB}}$,  $P_\text{c}$ and $\sigma_k^2$ are defined as in Lemma \ref{lemma2}, and $1/\text{SINR}_{kn}$ is given by (\ref{inverseSINR_MR}) and (\ref{inverseSINR_ZF}) for MR and ZF combining, respectively. The large-scale attenuation $\beta_{ljk}$ is defined as in Section \uppercase\expandafter{\romannumeral3} D. The operation $E_\mathbf{z}\{\cdot\}$ denotes taking the expectations over the UEs' random geographical distribution.
  \label{lemma3}
\end{lemma}

\begin{IEEEproof}
  Lemma \ref{lemma3} is a direct extension of Theorem \ref{theorem1}, where the SINR of each UE's TA is a random variable governed by the UE's location $\mathbf{z}_{jk}$, which is averaged out by the expectation operation $E_\mathbf{z}\{\cdot\}$.
\end{IEEEproof}

However, the expression in (\ref{Rrandom}) requires a large-scale Monte-Carlo simulation over the UEs' distributions, which becomes prohibitive when the number of UEs is large. Therefore we propose Theorem \ref{theorem2} to asymptotically lower-bound $R_k^\text{random}$ by $R_k^\text{random,LB}$ in a tractable form.

\begin{theorem}
At an asymptotically large number of BS antennas, the achievable uplink SE of UE $k$ in cell $0$ for random UE distribution is lower-bounded by:
\begin{equation}
\arraycolsep=1.0pt\def\arraystretch{1.3}
\begin{array}{rcl}
&& R_k^\text{random,LB} = \displaystyle\frac{T-B}{T} \left[\log_2\left(1 + N\chi^{-2}\right) + \log_2 N + \,\text{...} \right. \\
&& \left. P_\text{c} \log_2 P_\text{c} + (1-P_\text{c}) \log_2 \left( \displaystyle\frac{1 - P_\text{c}}{N-1} \right) \right] \,\, \left[\text{bits/s/Hz}\right],
\end{array}
\label{Rrandom_LB}
\end{equation}
where $P_\text{c}$ is given in (\ref{PcDef}) with $\sigma_k^2$ being replaced by $\chi^2$, and $\chi^2$ is given by (\ref{inverseSINR_MR_Aver}) and (\ref{inverseSINR_ZF_Aver}) for MR and ZF combining, respectively. Besides we have $\bar{\mu}_j^{(t)} \triangleq E_\mathbf{z} \{( \frac{\beta_{0jk}}{\beta_{jjk}})^t \}$.

\label{theorem2}
\end{theorem}

\begin{IEEEproof}
  The proof is provided in the Appendix.
\end{IEEEproof}

\begin{figure*}[t]
\normalsize
\begin{equation}
\chi_\text{MR}^2 = (1 + \epsilon_\text{s}) \displaystyle \sum_{j \in \Phi'}\bar{\mu}_j^{(2)} + \epsilon_\text{s} + \frac{N}{M} \left( \frac{\sigma_\text{N}^2}{\omega K P_\text{u}} + \sum_{j \in \Phi'_0} \bar{\mu}_j^{(1)} \right) \left( \frac{\sigma_\text{N}^2}{P_\text{u}} + K \sum_{j' \in \Phi_0} \bar{\mu}_{j'}^{(1)}\right).
\label{inverseSINR_MR_Aver}
\end{equation}

\begin{equation}
\chi_\text{ZF}^2 = \displaystyle\sum_{j\in\Phi'}\bar{\mu}_j^{(2)} + \displaystyle\frac{\sum_{n=1}^N r_n}{M-NK} \sum_{j\in\Phi'_0}\bar{\mu}_j^{(1)} \left( \sum_{j' \in \Phi'_0}\bar{\mu}_{j'}^{(1)} + (K-1)\theta_\omega \sum_{j' \in \Phi'_0} \bar{\mu}_{j'}^{(1)} + K \sum_{j'\notin\Phi'_0} \bar{\mu}_{j'}^{(1)} + \frac{\sigma_\text{N}^2}{P_\text{u}}  \right).
\label{inverseSINR_ZF_Aver}
\end{equation}

\hrulefill
\end{figure*}

Observe from (\ref{inverseSINR_MR_Aver}) and (\ref{inverseSINR_ZF_Aver}) that the SE expression is independent of the UE index $k$, since all the UEs share the same statistical features and power-control strategies. Moreover, the SE expression depends on the specific UE distribution via $\bar{\mu}_j^{(w)}$, which can be pre-computed based on a given UE distribution and then be applied to a system-level analysis in conjunction with the various parameters, such as $M$, $N$, $K$, etc. Similar to the conclusions drawn for fixed UE locations, we also have Corollary \ref{corollary2} for characterizing the large-scale behavior of the SE results given by Theorem \ref{theorem2}.

\begin{corollary}
The reciprocal of the SINR for MR and ZF combining derived for random UE locations converges to the following limits, when $M\rightarrow \infty$:
\begin{equation}
\arraycolsep=1.0pt\def\arraystretch{1.3}
\begin{array}{rcl}
\displaystyle \chi_\text{MR}^2 &\rightarrow& \displaystyle (1+\epsilon_\text{s}) \sum_{j \in \Phi'} \bar{\mu}_{j}^{(2)} + \epsilon_\text{s}, \\
\displaystyle \chi_\text{ZF}^2 &\rightarrow& \displaystyle \sum_{j \in \Phi'} \bar{\mu}_{j}^{(2)}.
\end{array}
\end{equation}
\label{corollary2}
\end{corollary}

Corollary \ref{corollary2} shows a similar result to Corollary \ref{corollary1}, hence the antenna arrangement strategy of (\ref{TranAntSpOpt}) is eminently suitable for performance optimization.

\subsection{Bound Tightness}
In this subsection, we explore the tightness of the asymptotic SE bounds proposed in Theorem \ref{theorem2} as verified by the simulation results obtained via Lemma \ref{lemma3}. We employ the system parameters as in Table \ref{TABLESimu} of Section \uppercase\expandafter{\romannumeral3} D. The $K$ UEs in each cell are assumed to be randomly located within the cell according to the \textit{uniform distribution}, while no UE is allowed to be closer to its serving BS than $r_\text{min}$. The simulation results of Lemma \ref{lemma3} and the coefficients $\bar{\mu}_j^{(w)}$ ($j \in \Phi_0$) are computed based on averaging $50,000$ random UE-location realizations.

\begin{figure}
\center{\includegraphics[width=0.95\linewidth]{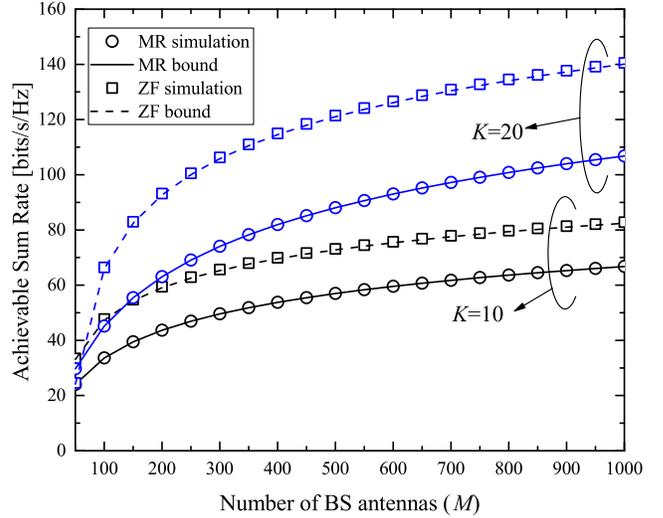}}
\caption{Achievable sum rate with respect to various $M$ and $K$. All the other parameters have been specified in Table \ref{TABLESimu}. The lines are computed based on Theorem \ref{theorem2} while the markers are obtained via Lemma \ref{lemma3}. }
\label{Fig_BoundTightnessRandom}
\end{figure}

In Fig.\ref{Fig_BoundTightnessRandom}, the achievable sum rates are depicted both for MR and ZF combining. The simulation results are seen to agree well with the theoretical bounds for all the $M$ values considered. Therefore we will use the analytical results of Theorem \ref{theorem2} in our following discourse. It is worth noting that, the bound presented in Fig.\ref{Fig_BoundTightnessRandom} is seen to be more accurate than the fixed UE location setting in Fig.\ref{Fig_BoundTightnessFixed}. The reason for this is that in Fig.\ref{Fig_BoundTightnessRandom} the simulation result is given by Lemma \ref{lemma3}, while Lemma \ref{lemma3} is actually generalized from Theorem \ref{theorem1}. Since the tightness of Theorem \ref{theorem1} has been validated by Fig.\ref{Fig_BoundTightnessFixed}, it is thus reasonable to use Lemma \ref{lemma3} as a benchmark in Fig.\ref{Fig_BoundTightnessRandom}. In short, in Fig.\ref{Fig_BoundTightnessRandom}, we are actually showing the accuracy between a tight lower bound and its ``secondary'' lower bound, hence the accuracy can be even better than that in Fig.\ref{Fig_BoundTightnessFixed}.

\subsection{Discussion on the information-theoretic capabilities of spatial modulation}

In recent years, many researchers have aimed for quantifying the information-theoretic capability of SM systems \cite{za2014mianalysis}-\cite{ibrahim2016achievable}. However, they cannot be directly applied in our framework. On the one hand, the channel model considered in our paper is different. As seen from (\ref{EqChnModel1}), due to the linear combining operations, the post-processing channel is an additive Gaussian noise channel, which has not been studied in previous research. On the other hand, as seen in Appendix \ref{AppC}, a very important property of our proposed bound in Lemma \ref{lemma2} is the convexity that we managed to prove rigorously. Thanks to this convex property, we can bypass the computationally exhaustive solution in Lemma \ref{lemma3}, and apply the more efficient asymptotic bound of Theorem \ref{lemma2}. Since the convex properties of the previous theoretical results have not yet been substantiated, their application is not recommended for verifying the simulation results, which thus prohibits their applications in a scenario of random UE locations.

\subsection{Discussion on the mathematical rigor of the asymptotic lower bound}
In this subsection we seek to provide a brief summarization on how we asymptotically lower-bound the uplink SE of massive SM-MIMOs.

In the setting of fixed UE locations, we firstly use Lemma \ref{lemma1} to quantify the post-processing SINR. Then we propose Lemma \ref{lemma2} to lower-bound the post-processing SE. In order to derive Theorem \ref{theorem1}, the SINR expression of Lemma \ref{lemma1} is applied both to MR and ZF combining at an asymptotically large $M$, where several large-$M$ approximations have been used during the derivation in Appendix \ref{AppB}, e.g. (\ref{HdaggHat_H0})$\sim$(\ref{E_g0kn_hjpkpnp_ZF1}). Note that the approximation of (\ref{E_g0kn_hjpkpnp_ZF1p5}) imposes a slight reduction of the effective SINR at an asymptotically high $M$, which also produces an asymptotic lower-bounding effect. The approximated SINR expressions (\ref{inverseSINR_MR}) and (\ref{inverseSINR_ZF}) are then combined with the SE lower bound of Lemma \ref{lemma2} to yield Theorem \ref{theorem1}. Therefore, Theorem \ref{theorem1} constitutes a lower bound when an asymptotically high $M$ is invoked, while the proposed expression is shown by Fig.\ref{Fig_BoundTightnessFixed} to represent a close approximation even when $M$ is limited.

In the context of random UE locations, Lemma \ref{lemma3} provides a computationally exhaustive solution to quantify the average SE, i.e. $R_k^\text{random} = E_\mathbf{z} \{R_k^\text{fixed,LB}\}$. Since $R_k^\text{fixed,LB}$ is an asymptotic bound according to Theorem \ref{theorem1}, $R_k^\text{random}$ is thus also an asymptotic lower bound. Theorem \ref{theorem2} succeeds in lower-bounding $R_k^\text{random}$ by exploiting the convexity of $R_k^\text{fixed,LB}$, in which some large-$M$ approximations are also utilized during the derivations of $\chi_\text{MR}^2$ and $\chi_\text{ZF}^2$. Therefore, the analytical results of Theorem \ref{theorem2}, i.e. $R_k^\text{random,LB}$, constitute a lower bound at asymptotically high $M$, while it is shown by Fig.\ref{Fig_BoundTightnessRandom} to exhibit a high accuracy even when $M$ is limited.

\section{Uplink SE Optimization}
In this section we seek to optimize the per-cell sum rate with respect to the number of UE antennas $N$. The sum rate is calculated based on Theorem \ref{theorem2}, while the simulation parameters are the same as Table \ref{TABLESimu} of Section \uppercase\expandafter{\romannumeral3} D unless stated otherwise.
\subsection{Impacts of Different $N$ on the Uplink SE}

\begin{figure}
\center{\includegraphics[width=0.95\linewidth]{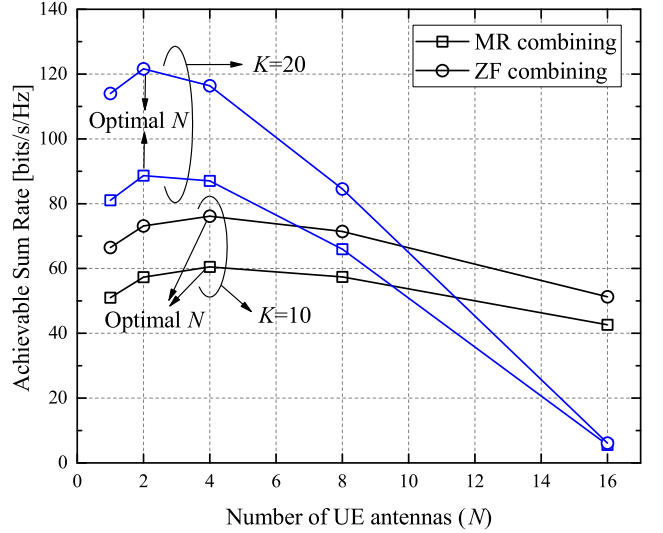}}
\caption{Achievable sum rates for $N\in \{1, 2, 4, 8, 16\}$ and $K \in \{10, 20\}$ when $D_\text{m} = 1$ m. All the other parameters have been specified in Table \ref{TABLESimu}.}
\label{Fig1_VariousN_K}
\end{figure}

We commence by exploring the impact of different values of $N$ on the uplink SE. The scenario of $N=1$ is also considered in our analysis, where we have modified the expression in Theorem \ref{theorem2} to:
\begin{equation}
\arraycolsep=1.0pt\def\arraystretch{1.3}
\begin{array}{rcl}
&& R_{k,N=1}^\text{random,LB} = \displaystyle\frac{T-B}{T} \log_2\left(1 + \chi^{-2}\right) \,\, \left[\text{bits/s/Hz}\right],
\end{array}
\label{Rk_N1}
\end{equation}
which only considers the single-input single-output (SISO) Shannon capacity and represents the uplink SE achieved by the conventional massive MIMO with single-antenna UEs. It is worth noting that we only consider specific $N$ values that are powers of two, which is required by the basic SM principle.


In Fig.\ref{Fig1_VariousN_K} we depicted the sum rates yielded by various $N$ values in conjunction with $K \in \{10, 20\}$ and $M = 512$, in which the optimal $N$ is $4$ and $2$ for $K=10$ and $20$, respectively. Observe from the figure that, for $K=20$, the sum rates of MR and ZF are drastically reduced, when $N$ increases from $2$ to $16$. The reason for this rapid SE degradation is that when $K$ is large, increasing $N$ leads to a significant reduction of the effective transmission ratio of $(T-B)/T$ (since $B = \omega N K$), which severely degrades the uplink SE.

\begin{figure}
\center{\includegraphics[width=0.95\linewidth]{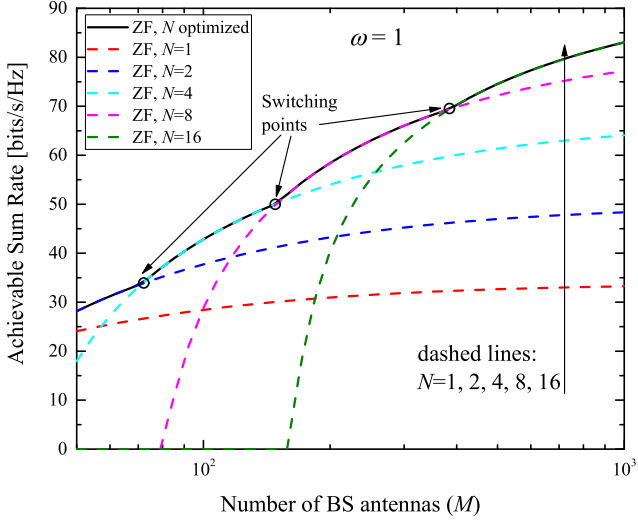}}
\caption{Achievable sum rates for $N\in \{1, 2, 4, 8, 16\}$ when $\omega=1$ and $D_\text{m} = 1$ m. All the other parameters have been specified in Table.\ref{TABLESimu}.}
\label{Fig1_VariousM}
\end{figure}

\begin{figure}
\center{\includegraphics[width=0.95\linewidth]{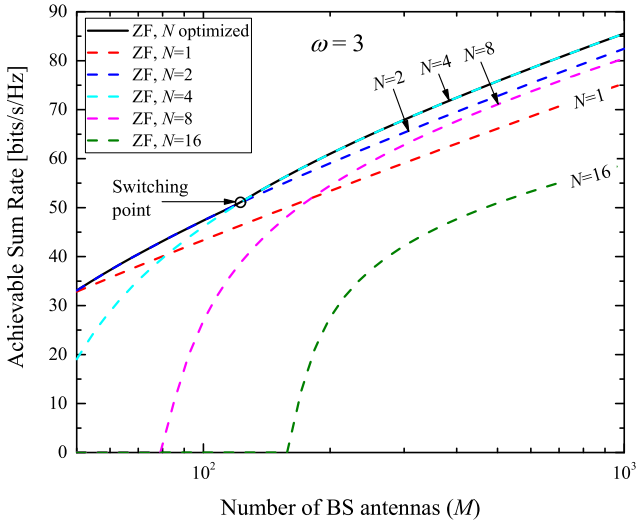}}
\caption{Achievable sum rates for $N\in \{1, 2, 4, 8, 16\}$ when $\omega=3$ and $D_\text{m} = 1$ m. All the other parameters have been specified in Table.\ref{TABLESimu}.}
\label{Fig1_VariousM_Alpha3}
\end{figure}

In Fig.\ref{Fig1_VariousM} and Fig.\ref{Fig1_VariousM_Alpha3} we quantify the uplink SE yielded by ZF combining for various $N$ values with dashed lines when we have $\omega=1$ and $\omega=3$, respectively. The $N$-optimized SE is also depicted with solid lines. For the case of $\omega=1$ in Fig.\ref{Fig1_VariousM}, the optimal $N$ is approximately $2$, $4$, $8$ and $16$ for $M < 70$, $70 \le M < 140$, $140 \le M < 400$ and $M \ge 400$, respectively. In the case of $\omega=3$, the optimal $N$ becomes $2$ and $4$ for $M < 120$ and $M \ge 120$, respectively. These examples encourage us to carefully select $N$ for the different values of $M$ and $\omega$, so that an uplink SE gain can indeed be achieved.

\subsection{Optimal $N$ for Various System Parameters}

\begin{figure}
\center{\includegraphics[width=0.95\linewidth]{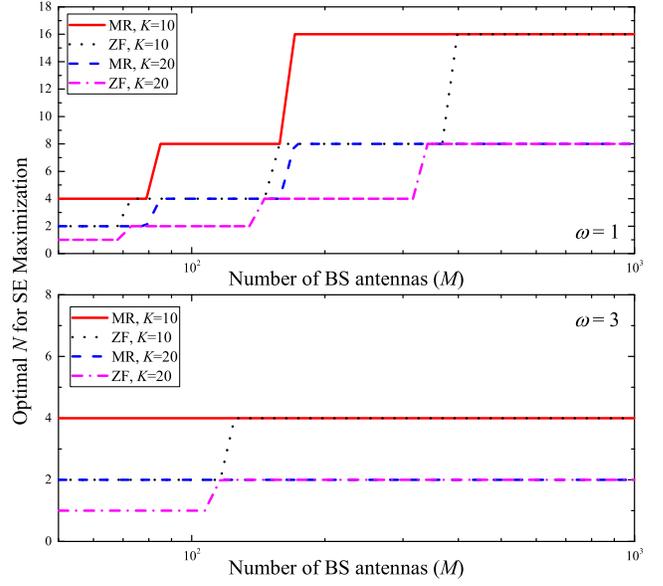}}
\caption{Optimal $N$ yielded by various $M$ values for $\omega \in \{1, 3\}$, $K \in \{10, 20\}$ and $D_\text{m} = 1$ m. All the other parameters have been specified in Table.\ref{TABLESimu}.}
\label{Fig2_OptimalN_M}
\end{figure}

Next we investigate the optimal $N$ values associated with various system parameters. For simplicity, we use $N^*$ to denote the optimal $N$ value. In Fig.\ref{Fig2_OptimalN_M}, $N^*$ yielded by various $M$ values associated with $\omega \in \{1, 3\}$ and $K \in \{10, 20\}$ are depicted. According to the figure, when $\omega = 1$, $N^*$ is seen to be increased upon increasing $M$ for both MR and ZF, while $N^*$ becomes less sensitive to $M$ for $\omega = 3$. Moreover, MR is observed to require a higher $N^*$ than ZF for the same $\omega$ and $K$, which substantiates our finding in the asymptotic analysis of Section \uppercase\expandafter{\romannumeral3} C namely that MR combining may benefit more from SM than ZF combining, when $M$ is limited.

\begin{figure}
\center{\includegraphics[width=0.95\linewidth]{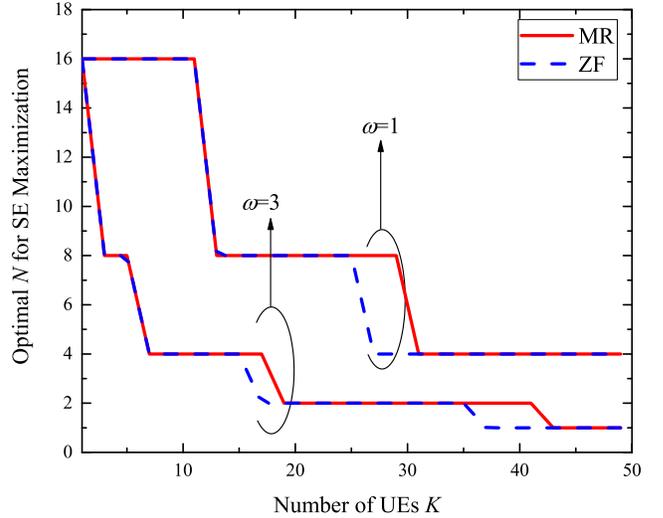}}
\caption{Optimal $N$ yielded by various $K$ for $\omega \in \{1, 3\}$ and $D_\text{m} = 1$ m. All the other parameters have been specified in Table.\ref{TABLESimu}.}
\label{Fig2_OptimalN_K}
\end{figure}

The relationship between $N^*$ and $K$ is more intuitively shown in Fig.\ref{Fig2_OptimalN_K}. According to the figure, upon increasing $K$ from $1$ to $50$, $N^*$ is gradually reduced to $4$ and $1$ for $\omega = 1$ and $3$, respectively. An intuitive explanation of this relationship between $N^*$ and $K$ is as follows. Using a larger $K$, the reduction of the normalized transmission time $(T-B)/T = 1-\omega NK/T$ when $N$ is increased becomes more severe, which rapidly neutralizes the benefits of increasing $N$. Hence $N^*$ becomes smaller when $K$ is increased. Moreover, Fig.\ref{Fig2_OptimalN_K} also shows that a higher level of inter-cell interference, i.e. a lower pilot reuse factor $\omega$ promotes the application of high-$N$ based massive SM-MIMOs. To be specific, when $K = 10$, the optimal $N^*$ is $16$ and $4$ for $\omega=1$ and $\omega=3$, respectively, while $N^*$ becomes $8$ and $2$ for $\omega=1$ and $\omega=3$ when $K=20$. Hence massive SM-MIMOs are more beneficial, when a higher level of inter-cell-interference is present.

\begin{figure}
\center{\includegraphics[width=0.95\linewidth]{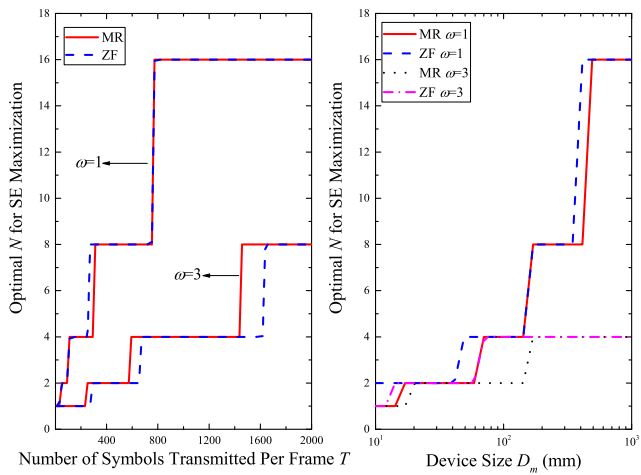}}
\caption{Optimal $N$ yielded by various $T$ (left) and $D_\text{m}$ (right) for $\omega \in \{1, 3\}$. All the other parameters have been specified in Table \ref{TABLESimu}.}
\label{Fig2_OptimalN_T_Dm}
\end{figure}

In Fig.\ref{Fig2_OptimalN_T_Dm}, the dependence of $N^*$ on $T$ and $D_\text{m}$ are also characterized, where $N^*$ is observed to be both increasing, when $T$ or $D_\text{m}$ is increased. The reduction of $D_\text{m}$ tends to increase the TA correlation for both MR and ZF combining, hence a smaller $N$ should be applied, when $D_\text{m}$ is small. Moreover, with $T$ increasing from $10$ to $2000$, the ratio $(T-\omega NK)/T$ tends to decrease much slower upon increasing $N$, and the SE gain achieved by employing SM thus becomes dominant, which leads to a higher $N^*$, as shown in Fig.\ref{Fig2_OptimalN_T_Dm}.


\subsection{Impact of the System Parameters on the $N$-optimized Uplink SE}
Let us now explore how the system parameters affect the achievable uplink SE with optimization over $N$. In the following simulation results we also provide the non-optimized SE associated with $N=1$ for benchmarking, which represents the uplink SE achieved by conventional massive MIMO systems relying on single-TA UEs.

\begin{figure}
\center{\includegraphics[width=0.95\linewidth]{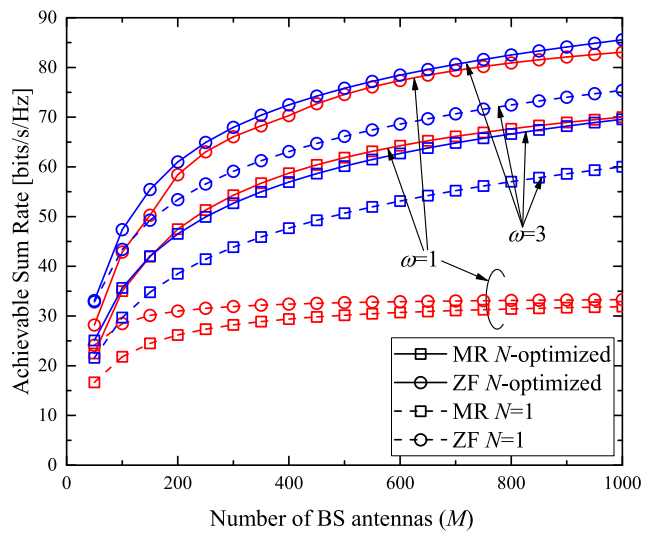}}
\caption{$N$-optimized per-cell SE with various $M$ and $\omega$ for $D_\text{m} = 1$ m (solid lines). The SE yielded by conventional massive MIMO with single-TA UEs is also provided by the dashed lines. All the other parameters have been specified in Table \ref{TABLESimu}.}
\label{Fig3_Noptimized_SE_MAlpha}
\end{figure}

We commence by exploring the impact of both $M$ and of the pilot reuse factor $\omega$ on the uplink SE in Fig.\ref{Fig3_Noptimized_SE_MAlpha}. For the SE associated with $N=1$, it is seen that $\omega=3$ yields a much higher SE than $\omega=1$, which is mainly due to the reduced inter-cell interference imposed by the high pilot reuse factor. For the $N$-optimized per-cell SE represented by solid lines, however, it is observed that the SE exhibits only modest variation when $\omega$ changes from $1$ to $3$, which suggests that the high inter-cell interference imposed by the less aggressive pilot reuse ($\omega=1$) has been compensated by employing massive SM-MIMOs.


\begin{figure}
\center{\includegraphics[width=0.95\linewidth]{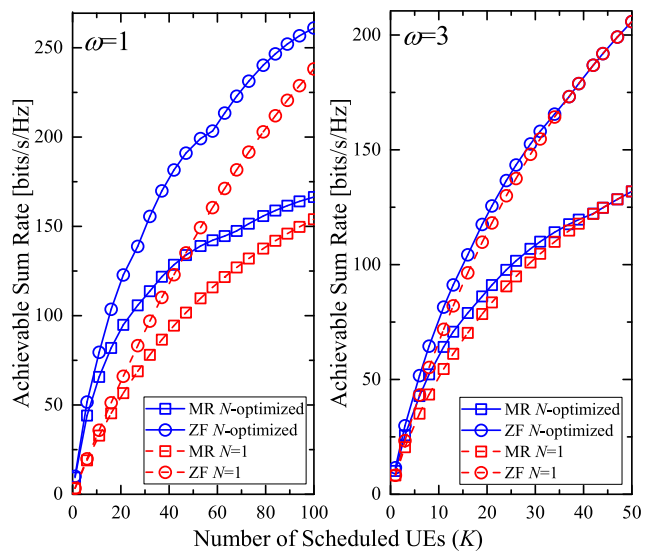}}
\caption{$N$-optimized per-cell SE with various $K$ and $\omega$ for $D_\text{m} = 1$ m (solid lines). The SE yielded by conventional massive MIMO with single-TA UEs is also provided by the dashed lines. All the other parameters have been specified in Table \ref{TABLESimu}.}
\label{Fig3_Noptimized_SE_K}
\end{figure}

\begin{figure}
\center{\includegraphics[width=0.95\linewidth]{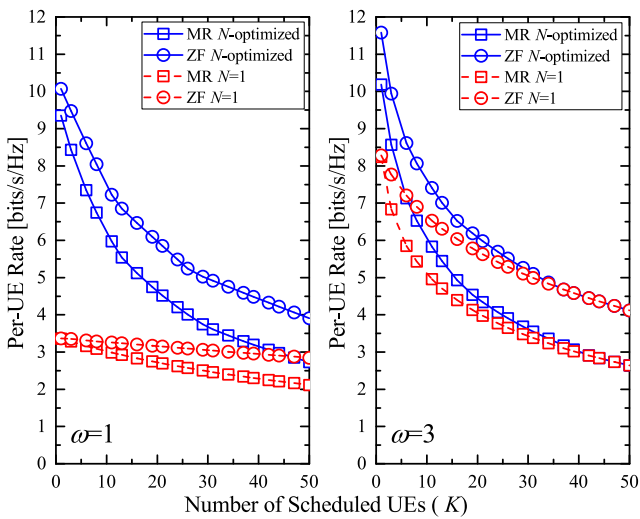}}
\caption{$N$-optimized per-UE SE with various $K$ and $\omega$ for $D_\text{m} = 1$ m (solid lines). The SE yielded by conventional massive MIMO with single-TA UEs is also provided by the dashed lines. All the other parameters have been specified in Table \ref{TABLESimu}.}
\label{Fig4_Noptimized_PU_SE}
\end{figure}

Next, the impact of $K$ is explored in Fig.\ref{Fig3_Noptimized_SE_K} both in conjunction with $\omega=1$ (left) and $3$ (right). In the case of $\omega = 1$, a significant performance gain can be harnessed by optimizing $N$, as shown in the left of Fig.\ref{Fig3_Noptimized_SE_K}. In the case of $\omega = 3$, however, the performance gain becomes lower and can only be achieved when $K$ is less than $40$. Both the plots of Fig.\ref{Fig3_Noptimized_SE_K} have revealed that massive SM-MIMOs combined with $N$-optimization are capable of outperforming the conventional massive MIMOs, when a limited number of UEs are being served. In Fig.\ref{Fig4_Noptimized_PU_SE}, the impact of $N$-optimization on the per-UE rate is also shown with respect to various $K$ values. It can be more explicitly seen that the per-UE rate is significantly improved by $N$-optimization when $K$ is limited, while the performance gain becomes lower for a large $K$, especially when $\omega = 3$. It is worth noting that due to the application of user-specific power control, the SE expression proposed in Theorem \ref{theorem2} is independent of the UE index $k$, which leads to a uniform rate distribution for all the UEs.


\begin{figure}
\center{\includegraphics[width=0.95\linewidth]{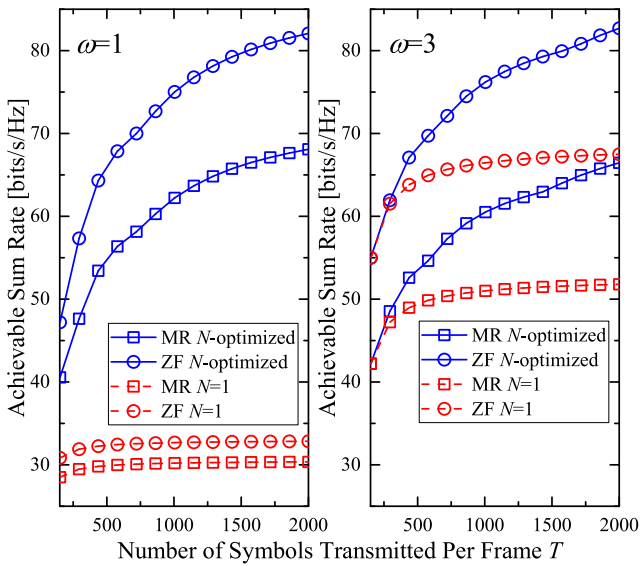}}
\caption{$N$-optimized per-cell SE with various $T$ and $\omega$ for $D_\text{m} = 1$ m (solid lines). The SE yielded by conventional massive MIMO with single-TA UEs is also provided by the dashed lines. All the other parameters have been specified in Table \ref{TABLESimu}.}
\label{Fig3_Noptimized_SE_T}
\end{figure}

\begin{figure}
\center{\includegraphics[width=0.95\linewidth]{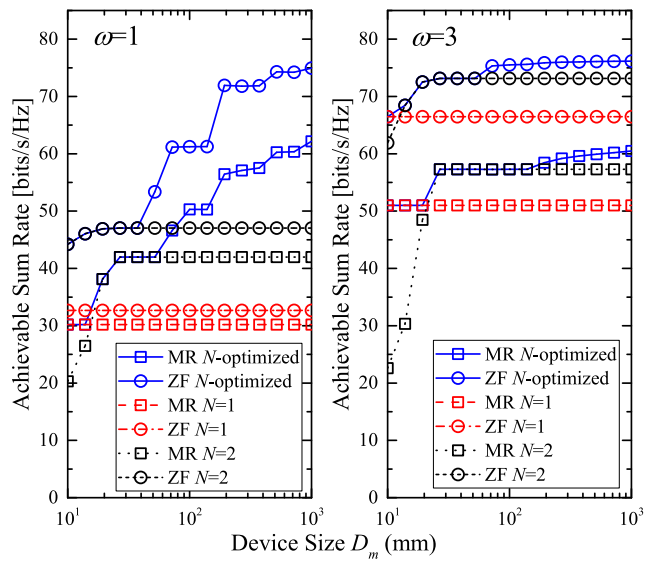}}
\caption{$N$-optimized per-cell SE with various $D_\text{m}$ and $\omega$ (solid lines). The SE yielded with $N=1$ and $N=2$ is also provided by the dashed lines and dotted lines, respectively. All the other parameters have been specified in Table \ref{TABLESimu}.}
\label{Fig3_Noptimized_SE_Dm}
\end{figure}

Finally, we explore the impact of $T$ and $D_\text{m}$ in Fig.\ref{Fig3_Noptimized_SE_T} and Fig.\ref{Fig3_Noptimized_SE_Dm}, respectively. As seen in Fig.\ref{Fig3_Noptimized_SE_T}, when $\omega=1$, an SE gain can be achieved by massive SM-MIMOs for the whole range of $T$ values considered. Moreover, according to Fig.\ref{Fig3_Noptimized_SE_Dm}, the performance gain is increased when $D_\text{m}$ is increased, while $N=1$ is only the optimal choice on condition when $D_\text{m}$ is very small.


\section{Conclusions and Future Research}
We investigated the uplink SE of a multi-cell massive SM-MIMO system relying on linear combining schemes. Asymptotic SE bounds with tractable formulas were derived, which eliminated the potentially prohibitive numerical simulations required for evaluating the achievable SE. The new SE expressions facilitated our novel system-level analysis, in which we maximized the per-cell SE by optimizing the number of UE antennas, and the resultant performance gain over the conventional massive MIMOs was substantiated via simulations. The optimal $N^*$ was found to be dependent on the various system parameters. By means of optimization over $N$, massive SM-MIMO was shown to be capable of outperforming the conventional massive MIMO for single-TA UEs.

In our future research, the massive SM-MIMO concept may be integrated with millimeter-wave (mmWave) systems \cite{bai2014mmWave} to accommodate more TAs in a single UE. By increasing the carrier's frequency to $20 \sim 60$ GHz, the TAs' spatial correlation can be reduced, hence facilitating for the uplink transmission to benefit more from employing SM schemes.


\appendices
\section{Proof of Lemma \ref{lemma2}}\label{AppA}
\begin{IEEEproof}
We commence the proof by showing that the mutual information term $I(\mathbf{y}_{\text{post}, k}; \mathbf{x}_{\text{SM}, k})$ is lower-bounded by $I(\mathbf{y}_{\text{AWGN}, k}; \mathbf{x}_{\text{SM}, k})$, with $\mathbf{y}_{\text{AWGN}, k}$ given by:
\begin{equation}
  \mathbf{y}_{\text{AWGN}, k} = \mathbf{x}_{\text{SM}, k} + \mathbf{w}_{\text{AWGN}, k},
\end{equation}
where the additive noise can be modeled as $\mathbf{w}_{\text{AWGN}, k} \sim \mathcal{CN}(\mathbf{0}, \sigma_k^2 \mathbf{I}_N)$ and $\sigma_k^2 = 1/N \sum_{n=1}^N \text{SINR}_{kn}^{-1}$. This can be verified by showing that $\mathbf{w}_{\text{AWGN}, k}$ has a higher differential entropy than $\mathbf{w}_{\text{eff}, k}$, hence resulting in a more grave reduction of the effective mutual information. The differential entropy of $\mathbf{w}_{\text{eff}, k}$ is given by:
\begin{equation}
  H(\mathbf{w}_{\text{eff}, k}) = \displaystyle \sum_{n=1}^N \log_2 \left(\pi e \text{SINR}_{kn}^{-1}\right) \le N \log_2(\pi e \sigma_k^2),
\end{equation}
of which the second inequality holds due to the concavity of $\log_2(\cdot)$. Since $H(\mathbf{w}_{\text{AWGN},k}) = N \log_2(\pi e \sigma_k^2)$, the inequality $I(\mathbf{y}_{\text{post}, k}; \mathbf{x}_{\text{SM}, k}) \ge I(\mathbf{y}_{\text{AWGN}, k}; \mathbf{x}_{\text{SM}, k})$ is hence proved.


According to the SM principle, $\mathbf{x}_{\text{SM}, k}$ can be expressed as $\mathbf{x}_{\text{SM}, k} = s_k \mathbf{a}_k$, where $s_k \sim \mathcal{CN}(0, N)$ is the Gaussian input, and $\mathbf{a}_k \in \{0, 1\}^N$ denotes the TA activity pattern. $I(\mathbf{y}_{\text{AWGN}, k}; \mathbf{x}_{\text{SM}, k})$ can hence be decomposed as:
\begin{equation}
  \arraycolsep=1.0pt\def\arraystretch{1.3}
  \begin{array}{rcl}
  && I(\mathbf{y}_{\text{AWGN}, k}; \mathbf{x}_{\text{SM}, k}) \\
  &=& I(\mathbf{y}_{\text{AWGN}, k}; \mathbf{a}_k) + I(\mathbf{y}_{\text{AWGN}, k}; s_k \vert \mathbf{a}_k),
  \end{array}
  \label{MutualInfDecomp}
\end{equation}
where $I(\mathbf{y}_{\text{AWGN}, k}; s_k \vert \mathbf{a}_k)$ can be quantified by Shannon's continuous-input continuous-output memoryless channel's (CCMC) capacity \cite{za2014mianalysis}, i.e $I(\mathbf{y}_{\text{AWGN}, k}; s_k \vert \mathbf{a}_k) = \log_2(1 + N \sigma_k^{-2})$.

The mutual information $I(\mathbf{y}_{\text{AWGN}, k}; \mathbf{a}_k)$ lacks a tractable formula, hence we propose to lower-bound it by $I(\hat{\mathbf{a}}_k ; \mathbf{a}_k)$, where $\hat{\mathbf{a}}_k$ is given by
\begin{equation}
  \hat{\mathbf{a}}_k = \displaystyle \mathbf{e}_{\displaystyle \arg\max_{1 \le t \le N} \left| \mathbf{y}_{\text{AWGN}, k} (t)\right|^2},
  \label{DetectionStrategy}
\end{equation}
where $\mathbf{e}_n$ represents the $n$-th column of an identity matrix $\mathbf{I}_N$. Due to the data-processing inequality in \cite[Theorem 2.8.1]{cover2006elements}, we have $I(\hat{\mathbf{a}}_k ; \mathbf{a}_k) \le I(\mathbf{y}_{\text{AWGN}, k}; \mathbf{a}_k)$, i.e. the detection of the TA imposes a mutual information loss. $I(\hat{\mathbf{a}}_k ; \mathbf{a}_k)$ can be formulated as:
\begin{equation}
  \arraycolsep=1.0pt\def\arraystretch{1.3}
  \begin{array}{rcl}
  && I(\hat{\mathbf{a}}_k ; \mathbf{a}_k) = \displaystyle \sum_{m=1}^N \sum_{n=1}^N \,\,\text{....} \\
  && \displaystyle \frac{\mathcal{P}(\hat{\mathbf{a}}_k = \mathbf{e}_n \vert \mathbf{a}_k = \mathbf{e}_m)}{N} \log_2 \frac{\mathcal{P}(\hat{\mathbf{a}}_k = \mathbf{e}_n \vert \mathbf{a}_k = \mathbf{e}_m)}{\mathcal{P}(\hat{\mathbf{a}}_k = \mathbf{e}_n)},
  \end{array}
  \label{IAA}
\end{equation}
where $\mathcal{P}(\hat{\mathbf{a}}_k = \mathbf{e}_n \vert \mathbf{a}_k = \mathbf{e}_m)$ denotes the probability of detecting TA $n$ as the active TA while TA $m$ is actually activated. Since the channel between $\mathbf{y}_{\text{AWGN}, k}$ and $\mathbf{x}_{\text{SM}, k}$ is simply AWGN, $\mathcal{P}(\hat{\mathbf{a}}_k = \mathbf{e}_n \vert \mathbf{a}_k = \mathbf{e}_m)$ can be simply characterized as:
\begin{equation}
  \mathcal{P}(\hat{\mathbf{a}}_k = \mathbf{e}_n \vert \mathbf{a}_k = \mathbf{e}_m) =
  \begin{cases}
    P_\text{c} & \text{if}\,\, m=n, \\
    \displaystyle\frac{1-P_\text{c}}{N-1}\,\, & \text{otherwise},
  \end{cases}
  \label{Pnm}
\end{equation}
where $P_\text{c}$ quantifies the correct TA detection probability and will be derived as follows. Given that TA $n$ is activated, $|\mathbf{y}_{\text{AWGN}, k}(t)|^2$ is distributed as:
\begin{equation}
  \displaystyle\frac{2|\mathbf{y}_{\text{AWGN}}(t)|^2}{N \delta_{t, n} + \sigma_k^2} \sim \mathcal{X}^2(2),
  \label{y_awgn_t}
\end{equation}
where $\delta_{t, n}$ equals $1$ and $0$ when $t=n$ and $t\ne n$, respectively. Therefore, the PDF of $\max_{t \ne n} |\mathbf{y}_{\text{AWGN}, k}(t)|^2$ is given by \cite{david2003order}:
\begin{equation}
\arraycolsep=1.0pt\def\arraystretch{1.3}
  \begin{array}{rcl}
  && \mathcal{P}(\displaystyle \max_{t \ne n} |\mathbf{y}_{\text{AWGN}, k}(t)|^2 = u) =\,\text{...}\\
  && \displaystyle \frac{D(N-1)}{2}(1 - e^{-\frac{u}{\sigma_k^2}})^{N-2} e^{-\frac{u}{\sigma_k^2}},
  \end{array}
  \label{Ap1_Eq1}
\end{equation}
where $D$ is the normalizing factor, of which the reciprocal is given as:
\begin{equation}
\arraycolsep=1.0pt\def\arraystretch{1.3}
  \begin{array}{rcl}
  && D^{-1} = \displaystyle \frac{N-1}{2} \int_{0}^\infty (1 - e^{-\frac{u}{\sigma_k^2}})^{N-2} e^{-\frac{u}{\sigma_k^2}} du\\
  &=& \displaystyle \frac{\sigma_k^2 (N-1)}{2} \sum_{r=0}^{N-2} \binom{N-2}{r} (-1)^r (1+r)^{-1},
  \end{array}
  \label{DValue}
\end{equation}
where the second equality is obtained by applying the binomial theorem. According to the detection strategy in (\ref{DetectionStrategy}), a TA detection event is successful if and only if $\mathbf{y}_{\text{AWGN}, k}(n)$ yields the maximal squared amplitude, i.e.
\begin{equation}
\arraycolsep=1.0pt\def\arraystretch{1.5}
  \begin{array}{rcl}
  && P_\text{c} = \mathcal{P}(|\mathbf{y}_{\text{AWGN}, k}(n)|^2 > \displaystyle \max_{t \ne n} |\mathbf{y}_{\text{AWGN}, k}(t)|^2) \\
  &=& \displaystyle\int_0^\infty \mathcal{P}(|\mathbf{y}_{\text{AWGN}, k}(n)|^2 > u) \mathcal{P}(\displaystyle \max_{t \ne n} |\mathbf{y}_{\text{AWGN}, k}(t)|^2 = u) du \\
  &=& \displaystyle \frac{D \sigma_k^2 (N-1)}{2} \sum_{r=0}^{N-2} \binom{N-2}{r} (-1)^r \displaystyle \left(r + \frac{N + 2\sigma_k^2}{N + \sigma_k^2}\right)^{-1},
  \end{array}
  \label{Ap1_Eq3}
\end{equation}
where $\mathcal{P}(|\mathbf{y}_\text{AWGN,k}(n)|^2 > u) = \exp[-u/(N+\sigma_k^2)]$ according to (\ref{y_awgn_t}). Replacing $D$ in (\ref{Ap1_Eq3}) by (\ref{DValue}), the expression of $P_\text{c}$ in (\ref{PcDef}) is hence proved. Finally, substituting (\ref{PcDef}) and (\ref{Pnm}) into (\ref{IAA}), the expression of $I(\hat{\mathbf{a}}_k ; \mathbf{a}_k)$ can be obtained, which completes the proof.
\end{IEEEproof}


\section{Proof of Theorem \ref{theorem1}}\label{AppB}
\begin{IEEEproof}
Firstly, we provide the proof for MR combining. According to (\ref{Esti_h_0kn}) and $\mathbf{g}_{0kn}^\text{MR} = \hat{\mathbf{h}}_{0kn}$, we have:
\begin{equation}
\arraycolsep=1.0pt\def\arraystretch{1.5}
\begin{array}{rcl}
  && \displaystyle E_\mathbf{h}\left\{  \left(\mathbf{g}_{0kn}^\text{MR}\right)^H \mathbf{h}_{0kn}\right\} = M \beta_{00k} \\
  && \displaystyle E_\mathbf{h}\left\{ \left\| \mathbf{g}_{0kn}^\text{MR} \right\|^2 \right\} = M \left( \displaystyle \sum_{j \in \Phi'_0} \frac{\beta_{00k}\beta_{0jk}}{\beta_{jjk}} + \frac{\sigma_\text{N}^2 \beta_{00k}}{\omega K P_\text{u}} \right).
\end{array}
\label{MRTerm1}
\end{equation}


As for $E_\mathbf{h}\{ | \left(\mathbf{g}_{0kn}^\text{MR} \right)^H \mathbf{h}_{j'k'n'} |^2\}$, we consider the following two cases:

1) $j' \in \Phi'_0$ and $k' = k$: based on the definition of $\mathbf{g}_{0kn}^\text{MR}$, the mean and covariance of $\mathbf{g}_{0kn}^\text{MR}$ conditioned on $\mathbf{h}_{j'kn'}$ can be formulated as:
\begin{equation}
\arraycolsep=1.0pt\def\arraystretch{1.5}
\begin{array}{rcl}
E\left\{ \mathbf{g}_{0kn}^\text{MR} \vert \mathbf{h}_{j'kn'} \right\} &=& \sqrt{\displaystyle \frac{\beta_{00k}}{\beta_{j'j'k}}} \epsilon_{n n'} \mathbf{h}_{j'kn'} \\
Cov\left\{ \mathbf{g}_{0kn}^\text{MR} \vert \mathbf{h}_{j'kn'} \right\} &=& \mathbf{G}_{0kn}^\text{MR} - \displaystyle \frac{\beta_{00k}\beta_{0j'k}}{\beta_{j'j'k}} \epsilon_{n, n'}^2 \mathbf{I}_M,
\end{array}
\label{g_MR_cond_ECov}
\end{equation}
where $\epsilon_{n, n'} = \mathbf{R}_\text{t}(n, n')$ and $\mathbf{G}_{0kn}^\text{MR} = (\sum_{j \in \Phi'_0} \frac{\beta_{00k} \beta_{0jk}}{\beta_{jjk}} + \frac{\sigma_\text{N}^2 \beta_{00k}}{\omega K P_\text{u}}) \mathbf{I}_M$. Therefore we have:
\begin{equation}
\arraycolsep=1.0pt\def\arraystretch{1.5}
\begin{array}{rcl}
&& E_\mathbf{h} \left\{ \left| \left(\mathbf{g}_{0kn}^\text{MR}\right)^H \mathbf{h}_{j'kn'} \right|^2 \right\} =  M\beta_{00k} \beta_{0j'k} \times \text{...}\\
&& \left( \displaystyle \frac{M \epsilon_{nn'}^2 \beta_{0j'k}}{\beta_{j'j'k}} +  \sum_{j \in \Phi'_0} \frac{\beta_{0jk}}{\beta_{jjk}} + \frac{\sigma_\text{N}^2}{\omega K P_\text{u}}  \right),
\end{array}
\label{MRTerm3_1}
\end{equation}
where the equality is yielded by applying the property of central complex-valued Wishart distribution given in \cite{tulino2004random}, i.e $E_\mathbf{h}\{ \| \mathbf{h}_{j'kn' }\|^4 \} = M(1+M) \beta_{0j'k}^2$.

2) $j' \notin \Phi'_0$ or $k' \ne k$: in this case, $\mathbf{g}_{0kn}^\text{MR}$ and $\mathbf{h}_{j'kn'}$ are independently distributed, hence we have:
\begin{equation}
  E_\mathbf{h} \left\{ \left| \left(\mathbf{g}_{0kn}^\text{MR} \right)^H \mathbf{h}_{j'k'n'} \right|^2 \right\} = \beta_{0j'k'} E_\mathbf{h} \left\{ \left\| \mathbf{g}_{0kn}^\text{MR} \right\|^2  \right\},
  \label{MRTerm3_2}
\end{equation}
where $E\{\|\mathbf{g}_{0kn}^\text{MR}\|^2\}$ has been formulated in (\ref{MRTerm1}). Substituting (\ref{MRTerm1}), (\ref{MRTerm3_1}) and (\ref{MRTerm3_2}) into (\ref{SINRBound}), the expression in (\ref{inverseSINR_MR}) can hence be obtained.

As for the case of ZF combining, according to (\ref{Hjk}), (\ref{Hj}), (\ref{Hhat0}) and (\ref{Aj}), $\hat{\mathbf{H}}_0$ can be equivalently formulated as $\hat{\mathbf{H}}_0 = \tilde{\mathbf{U}}_0 \tilde{\mathbf{G}}_0^{\frac{1}{2}}$, where $\tilde{\mathbf{U}}_0 \in \mathbb{C}_{M \times NK}$ is composed of i.i.d. $\mathcal{CN}(0, 1)$ elements, and $\tilde{\mathbf{G}}_0 = \text{diag} \{
  \sum_{j \in \Phi'_0} \frac{\beta_{0jk}\beta_{00k}}{\beta_{jjk}}\mathbf{R}_\text{t} + \frac{\sigma_\text{N}^2 \beta_{00k}}{\omega K P_\text{u}} \mathbf{I}_N\}_{k=1}^K$. Moreover, $E_\mathbf{h}\{\|\mathbf{g}_{0kn}^\text{ZF}\|^2\}$ equals the $[(k-1)N+n;  (k-1)N+n]$ element of the matrix $E_\mathbf{h}\{ \hat{\mathbf{H}}_0^\dagger (\hat{\mathbf{H}}_0^\dagger)^H \}$, which can be derived as follows:
\begin{equation}
\arraycolsep=1.0pt\def\arraystretch{1.5}
\begin{array}{rcl}
&& E_\mathbf{h}\left\{ \hat{\mathbf{H}}_0^\dagger \left( \hat{\mathbf{H}}_0^\dagger\right)^H \right\} = \displaystyle \frac{1}{\beta_{00k} (M - NK)} \times\,\,\text{...} \\
&& \text{diag}\left\{ \left( \displaystyle\sum_{j \in \Phi'_0} \frac{\beta_{0jk}}{\beta_{jjk}} \mathbf{R}_\text{t} + \frac{\sigma_\text{N}^2}{\omega K P_\text{u}} \mathbf{I}_N \right)^{-1} \right\}_{k=1}^K,
\end{array}
\label{E_Hdagg_Hdagg}
\end{equation}
in which the equality is obtained by applying $\hat{\mathbf{H}}_0 = \tilde{\mathbf{U}}_0 \tilde{\mathbf{G}}_0^{\frac{1}{2}}$ and $E_\mathbf{h}\{  (\tilde{\mathbf{U}}_0^H \tilde{\mathbf{U}}_0)^{-1} \} = (M-NK)^{-1}\mathbf{I}_{NK}$. In order to obtain a simplified expression, we assume that $\omega K P_\text{u} \gg 1$, which holds when the number of UEs or the effective SNR $P_\text{u} / \sigma_\text{N}^2$ is high. This then immediately yields $E_\mathbf{h}\{  \|\mathbf{g}_{0kn}^\text{ZF} \|^2 \}$ as:
\begin{equation}
  E_\mathbf{h}\left\{ \left\| \mathbf{g}_{0kn}^\text{ZF} \right\|^2 \right\} = \displaystyle \frac{r_n}{\beta_{00k}(M - NK)\left( \sum_{j \in \Phi'_0}\frac{\beta_{0jk}}{\beta_{jjk}} \right)},
  \label{E_g0kn_ZF}
\end{equation}
where $r_n$ is used to denote the $(n; n)$ element of $\mathbf{R}_\text{t}^{-1}$. As for $E_\mathbf{h}\{ (\mathbf{g}_{0kn}^\text{ZF})^H \mathbf{h}_{0kn} \}$, the following approximations are applied for a sufficiently large $M$ value:
\begin{equation}
\arraycolsep=1.0pt\def\arraystretch{1.5}
\begin{array}{rcl}
  && \hat{\mathbf{H}}_0^\dagger \mathbf{H}_0 \approx \\
  && \text{diag}\left\{ \left( \displaystyle \frac{\sigma_\text{N}^2}{\omega K P_\text{u}}\mathbf{R}_\text{t}^{-1} + \sum_{j \in \Phi'_0}\frac{\beta_{0jk}}{\beta_{jjk}} \mathbf{I}_N \right)^{-1}  \right\}_{k=1}^K.
\end{array}
\label{HdaggHat_H0}
\end{equation}

Since $E_\mathbf{h}\{(\mathbf{g}_{0kn}^\text{ZF})^H \mathbf{h}_{0kn}\}$ equals the $[(k-1)N+n; (k-1)N+n]$ element of $E_\mathbf{z}(\hat{\mathbf{H}}_0^\dagger \mathbf{H}_0)$, we thus have
\begin{equation}
  E_\mathbf{h}\left\{ \left(\mathbf{g}_{0kn}^\text{ZF}\right)^H \mathbf{h}_{0kn} \right\} \approx \left( \displaystyle \sum_{j \in \Phi'_0} \frac{\beta_{0jk}}{\beta_{jjk}}\right)^{-1},
  \label{E_g0kn_h0kn_ZF}
\end{equation}
where we have made the approximation $\omega K P_\text{u} \gg \sigma_\text{N}^2$.

Let us now derive $E_\mathbf{h} \{ |(\mathbf{g}_{0kn}^\text{ZF})^H \mathbf{h}_{j'k'n'}|^2\}$. Again we consider the following two cases:

1) $j' \in \Phi'_0$: the following approximation can be applied at a very large $M$ value:
\begin{equation}
\arraycolsep=1.0pt\def\arraystretch{1.5}
\begin{array}{rcl}
&& \hat{\mathbf{H}}_0^\dagger \mathbf{H}_{j'} \approx \text{diag}\left\{ \displaystyle \sqrt{\frac{\beta_{0j'k}^2}{\beta_{00k} \beta_{j'j'k}}} \times \,\,\text{...}\right. \\
&& \left. \left(\displaystyle  \frac{\sigma_\text{N}^2}{\omega K P_\text{u}}\mathbf{R}_\text{t}^{-1} + \sum_{j \in \Phi'_0} \frac{\beta_{0jk}}{\beta_{jjk}}\mathbf{I}_N \right)^{-1} \right\}_{k=1}^K.
\end{array}
\label{HdaggerHjpApprx}
\end{equation}

Hence we have
\begin{equation}
  (\mathbf{g}_{0kn}^\text{ZF})^H \mathbf{h}_{j'k'n'} \approx \beta_{0j'k} \mathbf{M}_k(n, n') \delta_{k, k'} / \sqrt{\beta_{00k}\beta_{j'j'k}},
  \nonumber
\end{equation}
where
\begin{equation}
  \mathbf{M}_k \triangleq \left( \frac{\sigma_\text{N}^2}{\omega K P_\text{u}} \mathbf{R}_\text{t}^{-1} + \sum_{j \in \Phi'_0}\frac{\beta_{0jk}}{\beta_{jjk}}\mathbf{I}_N \right)^{-1}.
  \nonumber
\end{equation}

Therefore the expectation term $E\{|(\mathbf{g}_{0kn}^\text{ZF})^H \mathbf{h}_{j'kn'}|^2\}$ can be approximately upper-bounded as follows when $M$ is sufficiently large:
\begin{equation}
\arraycolsep=1.0pt\def\arraystretch{2.5}
\begin{array}{rcl}
  && E_\mathbf{h}\left\{ \left| \left(\mathbf{g}_{0kn}^\text{ZF}\right)^H \mathbf{h}_{j'kn'} \right|^2\right\}  \lesssim \\
  && \displaystyle \frac{\beta_{0j'k}^2}{\beta_{00k}\beta_{j'j'k}} \mathbf{M}_{k}^2(n, n') + \beta_{0j'k}E\left\{ \left\|\mathbf{g}_{0kn}^\text{ZF}\right\|^2\right\}.
\end{array}
\label{E_g0kn_hjpkpnp_ZF1}
\end{equation}

For the case of $k' \ne k$, again we apply the independence assumption between $\mathbf{g}_{0kn}^\text{ZF}$ and $\mathbf{h}_{j'k'n'}$, which leads to:
\begin{equation}
E_\mathbf{h}\left\{ \left| \left(\mathbf{g}_{0kn}^\text{ZF}\right)^H \mathbf{h}_{j'k'n'} \right|^2\right\} \approx \theta_\omega \beta_{0j'k'} E_\mathbf{h} \left\{\left\|  \mathbf{g}_{0kn}^\text{ZF}\right\|^2  \right\},
\label{E_g0kn_hjpkpnp_ZF1p5}
\end{equation}
where $\theta_\omega < 1$ is a scaling factor allowing us to prevent the overestimation of $E\{|(\mathbf{g}_{0kn}^\text{ZF})^H \mathbf{h}_{j'k'n'}|^2\}$. Based on heuristic observations, by setting $\theta_\omega$ to $0.2$ and $0.01$, when $\omega=1$ and $\omega>1$ respectively, the SINR approximation tends to be relatively accurate.

2) $j' \notin \Phi'_0$:
in this case, $\mathbf{g}_{0kn}^\text{ZF}$ and $\mathbf{h}_{j'k'n'}$ are independently distributed, which yields:
\begin{equation}
  E_\mathbf{h}\left\{ \left| \left(\mathbf{g}_{0kn}^\text{ZF}\right)^H \mathbf{h}_{j'k'n'} \right|^2 \right\} = \beta_{0j'k'} E_\mathbf{h}\left\{\left\|  \mathbf{g}_{0kn}^\text{ZF}\right\|^2  \right\},
  \label{E_g0kn_hjpkpnp_ZF2}
\end{equation}
where $E\{\|\mathbf{g}_{0kn}^\text{ZF}\|^2\}$ has been given by (\ref{E_g0kn_ZF}). Finally, substituting (\ref{E_g0kn_ZF}), (\ref{E_g0kn_h0kn_ZF}), (\ref{E_g0kn_hjpkpnp_ZF1}), (\ref{E_g0kn_hjpkpnp_ZF1p5}) and (\ref{E_g0kn_hjpkpnp_ZF2}) into (\ref{SINRBound}), and applying the assumption that $\omega K P_\text{u}/\sigma_\text{N}^2 \gg 1$, the expression of $1/\text{SINR}_{kn}^\text{ZF}$ in (\ref{inverseSINR_ZF}) can hence be obtained.
\end{IEEEproof}

\section{Proof of Theorem \ref{theorem2}}\label{AppC}
\begin{figure*}[t]
\normalsize
\begin{equation}
\kappa_{kn}^\text{MR} = (1 + \epsilon_\text{s}) \displaystyle \sum_{j \in \Phi'}\bar{\mu}_j^{(2)} + \epsilon_\text{s} + \frac{N}{M} \left( \frac{\sigma_\text{N}^2}{\omega K P_\text{u}} + \sum_{j \in \Phi'_0} \bar{\mu}_j^{(1)} \right) \left( \frac{\sigma_\text{N}^2}{P_\text{u}} + K \sum_{j' \in \Phi_0} \bar{\mu}_{j'}^{(1)}\right) + \displaystyle \frac{N}{M} \sum_{j \in \Phi'} \text{var}(\mu_{jk}).
\label{inverseSINR_MR_Ap3_New}
\end{equation}

\begin{equation}
\arraycolsep=1.0pt\def\arraystretch{1.5}
\begin{array}{rcl}
\kappa_{kn}^\text{ZF} &=& \displaystyle\sum_{j\in\Phi'} \bar{\mu}_{j}^{(2)} + \displaystyle\frac{r_n N}{M-NK} \displaystyle\sum_{j\in\Phi'_0} \bar{\mu}_{j}^{(1)} \left( \sum_{j' \in \Phi'_0}\bar{\mu}_{j'}^{(1)} + (K-1)\theta_\omega \sum_{j' \in \Phi'_0} \bar{\mu}_{j'}^{(1)} + K \sum_{j'\notin\Phi'_0} \bar{\mu}_{j'}^{(1)} + \frac{\sigma_\text{N}^2}{P_\text{u}}  \right) + \text{...} \\
&& \displaystyle \frac{r_n N}{M - NK} \sum_{j \in \Phi'} \text{var}(\mu_{jk})
\label{inverseSINR_ZF_Ap3_New}
\end{array}
\end{equation}
\hrulefill
\end{figure*}

\begin{IEEEproof}
Firstly we will prove that $R_k^\text{fixed,LB}$ in Lemma \ref{lemma2} is \textit{convex} with respect to $\sigma_k^2$, which leads to a direct application of Jensen's inequality as follows:
\begin{equation}
  R_k^\text{random} = E_\mathbf{z}\left\{ R_k^\text{fixed,LB} \right\} \ge R_k^\text{random,LB},
  \label{Ap3_Eq0}
\end{equation}
in which $R_k^\text{random,LB}$ is obtained by replacing $\sigma_k^2$ in (\ref{Rfixed_LB}) with $E_\mathbf{z}\{\sigma_k^2\}$. We commence by formulating $R_k^\text{fixed,LB}$ as $R_k^\text{fixed,LB} = \frac{T-B}{T} \{f_1\left(\sigma_k^2\right) + f_2[g(\sigma_k^2)] \}$, where we have $f_1(x) = \log_2(1+N/x) + \log_2 N$, $f_2(x) = x\log_2 x + (1-x)\log_2[(1-x)/(N-1)]$, and $g(x)$ is yielded by substituting $\sigma_k^2$ in (\ref{PcDef}) with $x$. The convexity of $f_1(x)$ can hence be immediately validated. In order to demonstrate the convexity of $f_2[g(x)]$, we seek to prove the following three properties:

(1) Convexity of $f_2(x)$: The convexity of $f_2(x)$ can be explicitly proved according to the convexity of function $x \log_2 x + (1-x) \log_2(1-x)$ with respect to $x$.

(2) Convexity of $g(x)$: Aided with the expressions of $P_\text{c}$ and $D^{-1}$ in (\ref{Ap1_Eq3}) and (\ref{DValue}), it can be derived that:
\begin{equation}
g(x) = P_\text{c} \vert_{\sigma_k^2 = x} \propto \displaystyle \int_0^\infty \left(1-e^{-t}\right)^{N-2} e^{-2t} e^{\frac{Nt}{x + N}} dt.
\label{Ap3_Eq3}
\end{equation}


Therefore the convexity of $g(x)$ can be instantly substantiated according to the convexity of $e^{Nt/(x+N)}$, since $(1-e^{-t})^{N-2} e^{-2t} > 0$ also holds.


(3) Monotonic nature of $f_2(x)$: According to the definition of $P_\text{c}$ in (\ref{PcDef}), it can be concluded that we have $1/N \le g(x) \le 1$ with $x$ varying from $\infty$ to $0$. Therefore we only consider the monotonic nature of $f_2(x)$ for $1/N \le x \le 1$. Calculating the derivative of $f_2(x)$ yields:
\begin{equation}
f'_2(x) = \log_2 \displaystyle\frac{(N-1)x}{1-x} \ge \log_2 \displaystyle\frac{(N-1)\min\{x\}}{1-\min\{x\}} = 0,
\label{f2_derivative}
\end{equation}
where $\min\{x\} = 1/N$. Hence $f_2(x)$ is monotonically increasing with $x$ varying from $1/N$ to $1$.

Given that the above three properties have been proved, the convexity of $f_2[g(x)]$ can thus be proved according to the convexity of functions' compositions \cite{boyd2004convex}, hence $R_k^\text{random,LB}$ constitutes a solid lower bound for $R_k^\text{random}$. In order to calculate $R_k^\text{random,LB}$, $E_\mathbf{z}\{\sigma_k^2\} = 1/N \sum_{n=1}^N E_\mathbf{z}\{\text{SINR}_{kn}^{-1}\}$ must be obtained, hence we need to specify the values of $\kappa_{kn}^\text{MR} \triangleq E_\mathbf{z}\{1/\text{SINR}_{kn}^\text{MR}\}$ and $\kappa_{kn}^\text{ZF} \triangleq E_\mathbf{z}\{1/\text{SINR}_{kn}^\text{ZF}\}$, when a specific UEs' distribution is given. Based on (\ref{inverseSINR_MR}) and (\ref{inverseSINR_ZF}), the values of $\kappa_{kn}^\text{MR}$ and $\kappa_{kn}^\text{ZF}$ can hence be respectively given by (\ref{inverseSINR_MR_Ap3_New}) and (\ref{inverseSINR_ZF_Ap3_New}), where $\text{var}(\mu_{jk})$ represents the variance of $\mu_{jk}$. Note that the impact of the variance term $\sum_{j \in \Phi'}\text{var}(\mu_{jk})$ is asymptotically reduced to zero with the increase of $M$, we hence neglect the variance terms in $\kappa_{kn}^\text{MR}$ and $\kappa_{kn}^\text{ZF}$ at an asymptotically large $M$ for the sake of simplicity. Finally, by substituting $E_\mathbf{z}\{\sigma_k^2\}$ with $\chi_\text{MR}^2 = \frac{1}{N}\sum_{n=1}^N \kappa_{kn}^\text{MR}$ and $\chi_\text{ZF}^2 = \frac{1}{N}\sum_{n=1}^N \kappa_{kn}^\text{ZF}$, the asymptotic SINR's reciprocals in (\ref{inverseSINR_MR_Aver}) and (\ref{inverseSINR_ZF_Aver}) are thus obtained, which completes the proof.\end{IEEEproof}

\begin{IEEEbiography}[{\includegraphics[width=1in,height=1.25in,clip,keepaspectratio]{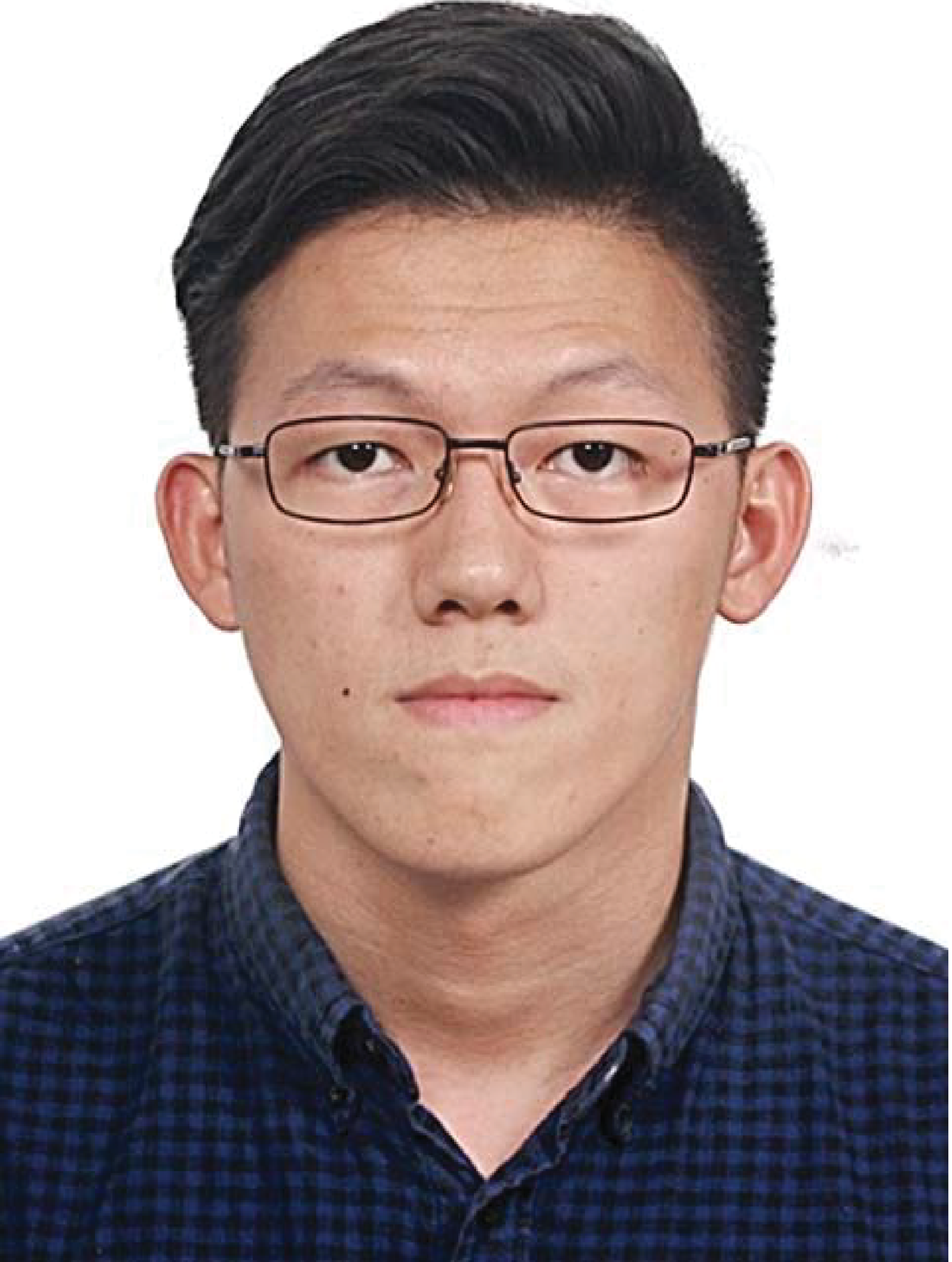}}]{Longzhuang He}
(S'15) received the bachelor's degree from the Department of Electronic Engineering at Tsinghua University, Beijing, China, in 2014. He is now pursuing the Ph.D. degree at the Department of Electronic Engineering in Tsinghua University.
His research interests include multiple-input multiple-output (MIMO) systems, spatial modulation (SM) technique, and communication signal processing.
\end{IEEEbiography}

\begin{IEEEbiography}[{\includegraphics[width=1in,height=1.25in,clip,keepaspectratio]{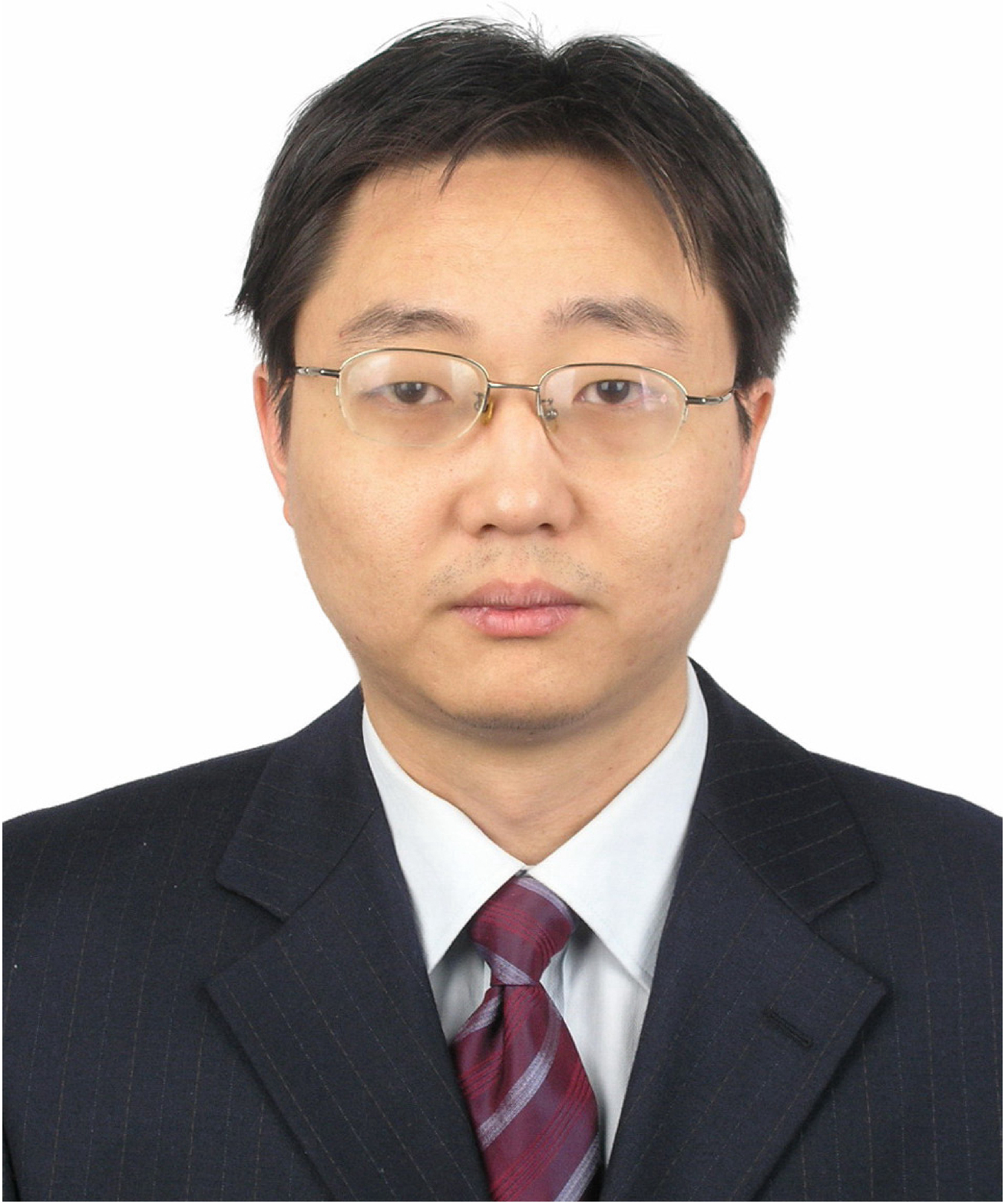}}]{Jintao Wang}
(SM'12) received the B.Eng. and Ph.D. degrees in electrical engineering both from Tsinghua University, Beijing, China, in 2001 and 2006, respectively. From 2006 to 2009, he was an Assistant Professor in the Department of Electronic Engineering at Tsinghua University. Since 2009, he has been an Associate Professor and Ph.D. Supervisor. He is the Standard Committee Member for the Chinese national digital terrestrial television broadcasting standard. His current research interests include space-time coding, MIMO, and OFDM systems. He has published more than 100 journal and conference papers and holds more than 40 national invention patents.
\end{IEEEbiography}

\begin{IEEEbiography}[{\includegraphics[width=1in,height=1.25in,clip,keepaspectratio]{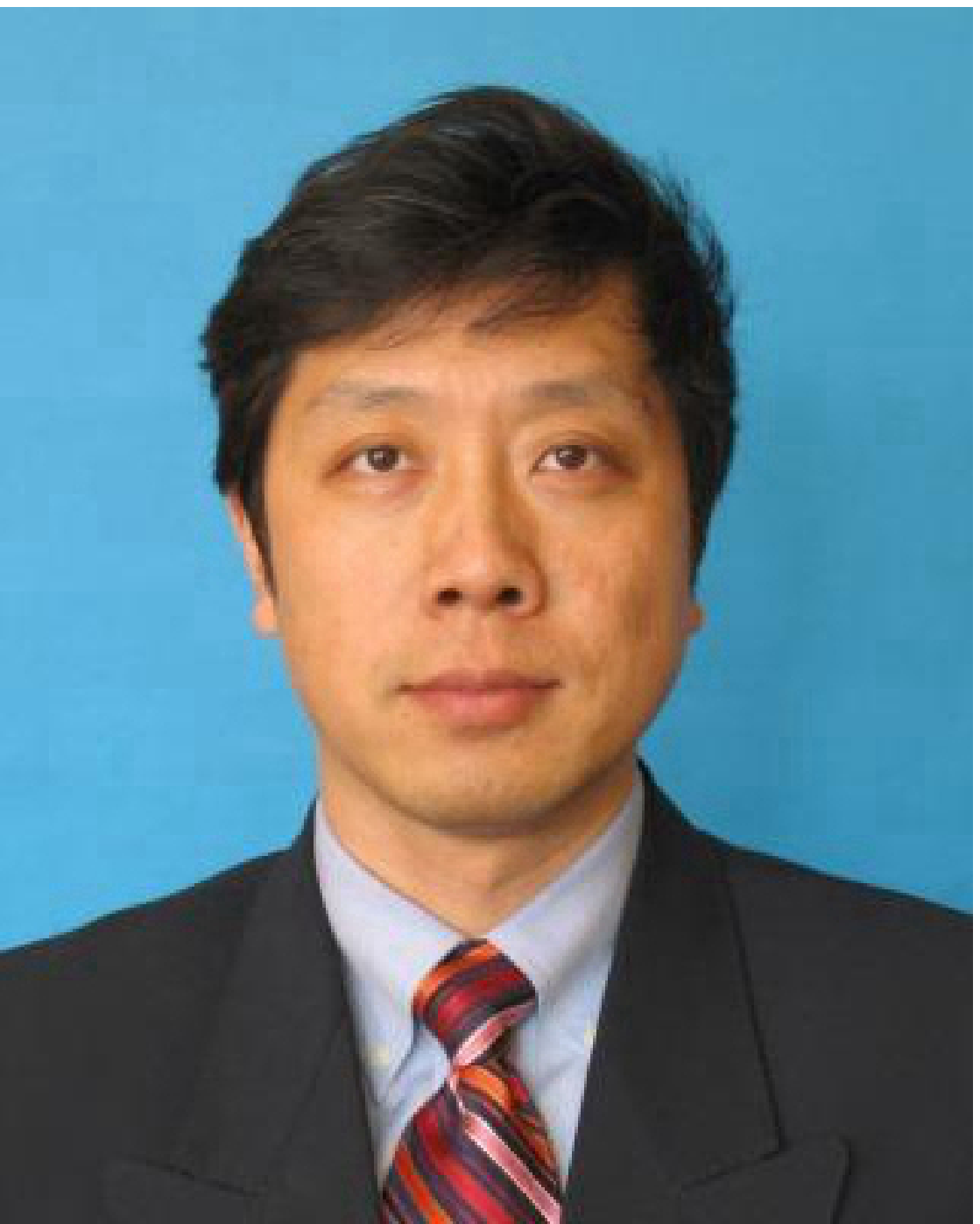}}]{Jian Song}
(M'06-SM'10-F'16) received the B.Eng and Ph.D. degrees in electrical engineering from Tsinghua University, Beijing, China, in 1990 and 1995, respectively. He worked for the same university upon his graduation and has worked at The Chinese University of Hong Kong and University of Waterloo, Canada, in 1996 and 1997, respectively. He has been with Hughes Network Systems in Germantown, Maryland, USA, for seven years before joining the faculty team at Tsinghua in 2005 as a Professor. Currently, he is the Director of Tsinghua DTV Technology R\&D Center. He has been working in quite different areas of fiber-optic, satellite and wireless communications, as well as the power-line communications. His current research interest is in the area of digital TV broadcasting. Dr. Song has published more than 110 peer-reviewed journal and conference papers. He holds two U.S. and more than 20 Chinese patents. He is a Fellow of IEEE and IET.
\end{IEEEbiography}

\begin{IEEEbiography}[{\includegraphics[width=1in,height=1.25in,clip,keepaspectratio]{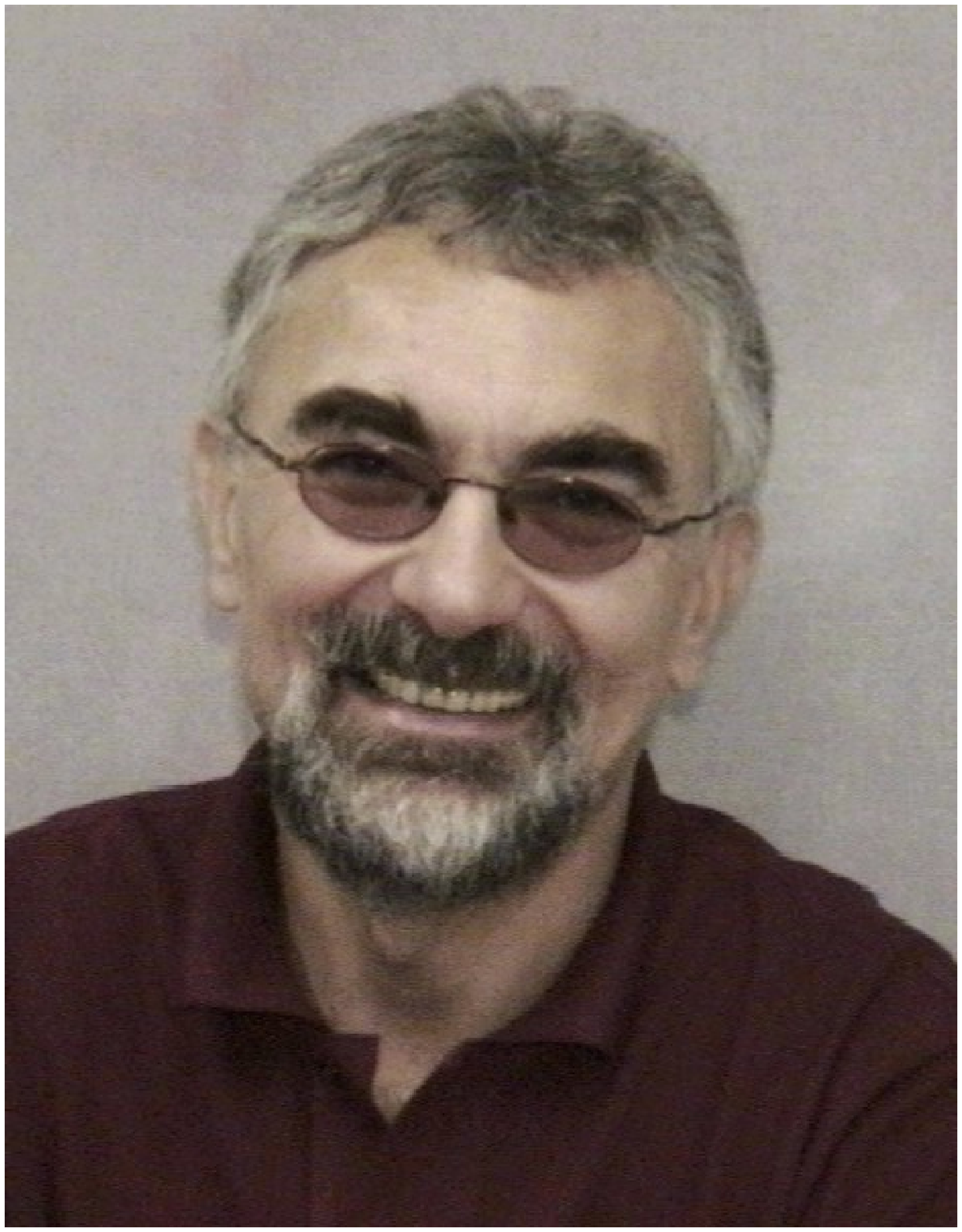}}]{Lajos Hanzo}
(http://www-mobile.ecs.soton.ac.uk) FREng, FIEEE, FIET, Fellow of EURASIP, DSc received his degree in electronics in 1976 and his doctorate in 1983. In 2009 he was awarded an honorary doctorate by the Technical University of Budapest and in 2015 by the University of Edinburgh. In 2016 he was admitted to the Hungarian Academy of Science. During his 40-year career in telecommunications he has held various research and academic posts in Hungary, Germany and the UK. Since 1986 he has been with the School of Electronics and Computer Science, University of Southampton, UK, where he holds the chair in telecommunications. He has successfully supervised 111 PhD students, co-authored 18 John Wiley/IEEE Press books on mobile radio communications totalling in excess of 10 000 pages, published 1600+ research contributions at IEEE Xplore, acted both as TPC and General Chair of IEEE conferences, presented keynote lectures and has been awarded a number of distinctions. Currently he is directing a 60-strong academic research team, working on a range of research projects in the field of wireless multimedia communications sponsored by industry, the Engineering and Physical Sciences Research Council (EPSRC) UK, the European Research Council's Advanced Fellow Grant and the Royal Society's Wolfson Research Merit Award. He is an enthusiastic supporter of industrial and academic liaison and he offers a range of industrial courses. He is also a Governor of the IEEE VTS. During 2008 - 2012 he was the Editor-in-Chief of the IEEE Press and a Chaired Professor also at Tsinghua University, Beijing. For further information on research in progress and associated publications please refer to http://www-mobile.ecs.soton.ac.uk. Lajos has 27 000+ citations and an H-index of 63.
\end{IEEEbiography}

\end{document}